\documentclass[12pt]{article}

\usepackage{epsfig}
\usepackage{amssymb}
\usepackage{bbm}
\usepackage{color}
\usepackage{ulem}

\usepackage{enumitem}
\usepackage{graphicx}
\usepackage{subfigure}
\usepackage{amsmath}
\usepackage{bbm}

\def\bi{\begin{itemize}}
\def\ei{\end{itemize}}
\def\bc{\begin{center}}
\def\ec{\end{center}}
\def\ben{\begin{enumerate}}
\def\een{\end{enumerate}}
\def\be{\begin{equation}}
\def\ee{\end{equation}}
\def\bea{\begin{eqnarray}}
\def\eea{\end{eqnarray}}

\def\hub{{\mathcal H}}

\begin{document}

\begin{center}
{\Large TASI 2009 - LECTURES ON COSMIC ACCELERATION}
\\[0.5cm]
{Rachel Bean}
\\[0.5cm]
{Department of Astronomy, Cornell University,\\
Ithaca, NY 14853, USA}
\end{center}
\vspace{0.3cm}
{ \noindent \textbf{Abstract} \\[0.3cm]
In this series of lectures we review observational evidence for, and theoretical investigations into, cosmic acceleration and dark energy. The notes are in four sections. First I review the basic cosmological formalism to describe the expansion history of the universe and how distance measures are defined. The second section covers the evidence for cosmic acceleration from cosmic distance measurements. Section 3 discusses the theoretical avenues being considered to explain the cosmological observations and section 4 discusses how the growth of inhomogeneities and large scale structure observations might help us pin down the theoretical origin of cosmic acceleration.
}



\section{Describing spacetime: distances and General Relativity}

Here we review the basics of General Relativity that allow us to describe the expansion history of a homogeneous and isotropic Universe -- a reasonable description of the Universe on the largest cosmic scales.

\subsection{Units}

First a note on units. Throughout, we will often set $c=k_B=\hbar=1$. This is just a change of units making time, distance and temperature consistent with units of energy,
\bea
\hbar \times time^{-1} &\leftrightarrow& \hbar c \times distance^{-1} \leftrightarrow k_B \times temp  \leftrightarrow  energy 
\eea

\noindent  It is often useful to use astrophysical, as well as SI, units of length. 
\bi
\item The {\bf Astronomical \ unit} is a good unit for solar system scale distances:
\bea
\mathrm{Astronomical \ unit (AU)} &=& \mathrm{Earth-Sun \ distance}
\\ & =& 1.49\times 10^{11}m
\\
\mathrm{Solar  \ system} &\sim&  40AU
\eea
\item Trigonometric parallax (how the apparent position of objects moves as the Earth orbits the Sun) defines the {\bf parsec} unit of length
\bea
\mathrm{Parsec (pc)} &=& \mathrm{Distance \ to \ object \ with  \ parallax \  of \ 1 \ arcsec}
\\
1 pc &=& \frac{1AU}{1arcsec}
\\ &=& 206,265 AU
\\ &=& 3.09 \times 10^{16}m
\eea
The $pc$ is useful unit to describe astrophysical distances 
\bea
\mathrm{Galactic \ radii} &\sim & 1-50 kpc
\\
\mathrm{Intergalactic \ scales} &\sim& 1 Mpc
\\
\mathrm{Galaxy \ cluster \ radii} &\sim& 1 Mpc
\\
\mathrm{Observable \ universe} &\sim& 4Gpc
\eea 
\ei

\subsection{Cosmic dynamics}
On scales above a few hundred Mpc the universe is well-approximated as homogeneous (no preferred location) and isotropic (no preferred direction). Being homogeneous and isotropic implies
\bi
\item  the universe can have one of three  types of geometry: negatively curved, positively curved or flat, with a radius of curvature, $R$, that is the same at every point in space.

\begin{figure}
\begin{center}
\includegraphics[width=5in,angle=0]{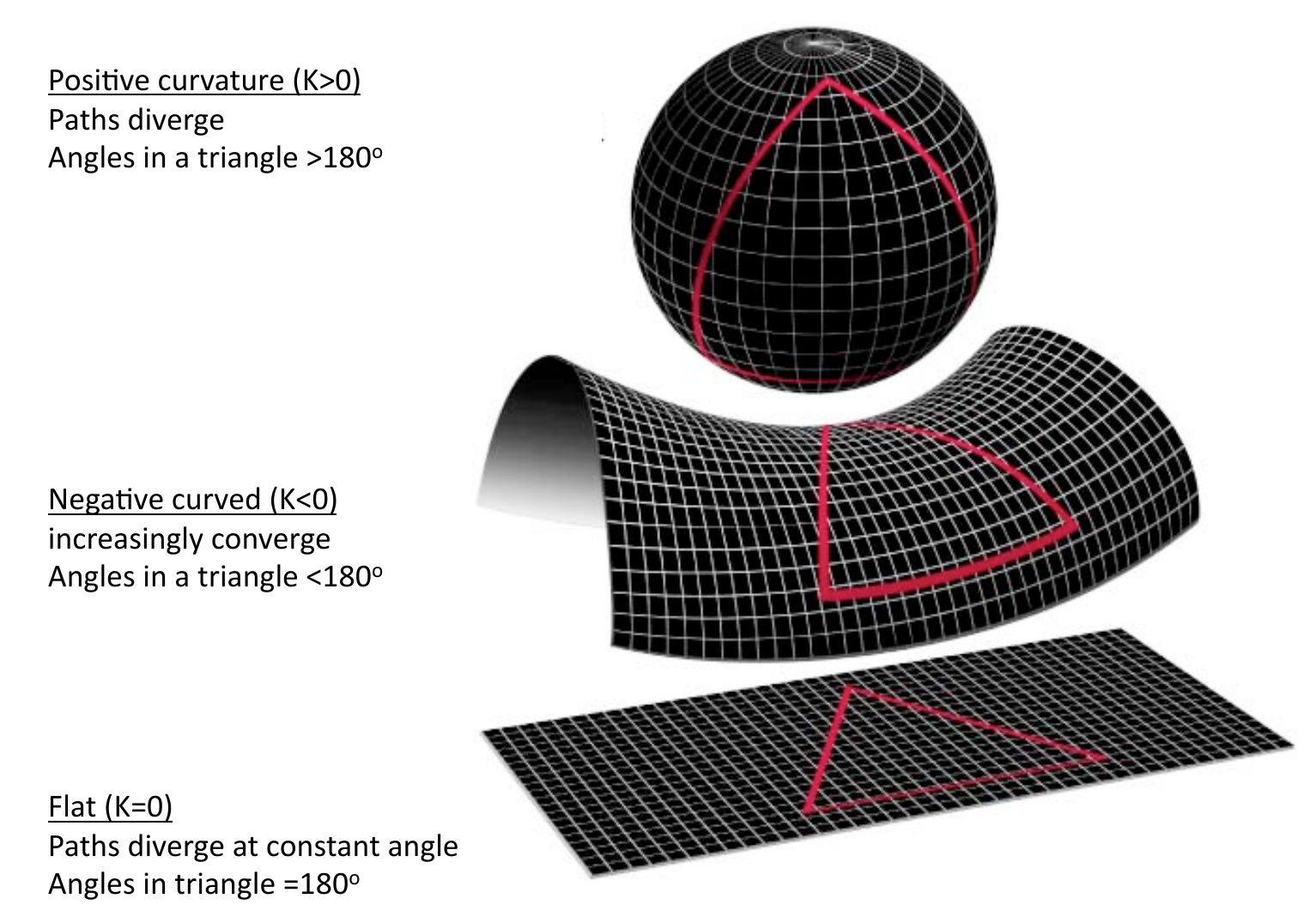}
\caption{The three types of curvature consistent with a homogeneous and isotropic space.}
\label{rb_fig1}
\end{center}
\end{figure}

\item the universe's expansion rate has to be the same at every place, i.e. it has only time, but no spatial, dependence,  described by a {\bf scale factor $a(t)$}. We will assume a convention where $a=1$ today.
 \ei

The relationship between distances and time intervals in a homogenous and isotropic universe can be conveniently described by the {\bf Friedmann Robertson Walker (FRW) metric}. This is commonly written in one of two, equivalent, forms:

Using spherical polar coordinates, $r$ is the comoving radial distance, $d\Omega = d\theta^2+\sin^2\theta d\phi^2$  and physical time $t$,
\bea
ds^2 &=& -dt^2 + a(t)^2\left[dr^2 + S_{\kappa}^2(r,R)d\Omega\right] \label{frw}
\eea
or with a change of variable $x\equiv S_{\kappa}(r,R)$, $K=\kappa/R^2$,
\bea
ds^2 &=& -dt^2 + a(t)^2\left[\frac{dx^2}{1-Kx^2} + x^2d\Omega\right]
\eea
where $\kappa$, $K\equiv \kappa /R^2$ and $S_\kappa(r,R)$ depend upon the geometry 
\bc
\begin{tabular}{|c|c|c|c|}
\hline
Geometry &  $ \kappa$ & $K = \kappa/R^{2}$&  $S_{\kappa}(r,R)$ 
\\ \hline
Flat &  $0$ &  0& $r$ 
\\ \hline
Closed &  $+1$ & $>0$& $R\sin\left(\frac{r}{R}\right)$
\\ \hline
Open &  -1 & $<0$& $R\sinh\left(\frac{r}{R}\right)$ 
\\ \hline
\end{tabular}
\ec
We can also define {\bf conformal time $d\tau\equiv dt/a(t)$} so that the FRW metric can be written
\bea
ds^2 &=& a(\tau)^2\left[-d\tau^2 + dr^2 + S_{\kappa}^2(r,R)d\Omega\right]
\eea

Finally we can write the space time element, $ds^2$, in a coordinate independent way, in terms of a {\bf metric}, $g_{\mu\nu}$ for a generic 4D space-time coordinate system $x^{\mu}$ where $\mu=0,1,2,3$. $\mu=0$ is reserved for the time coordinate and $\mu=1,2,3$ are for the 3 spatial coordinates. If we are just referring to the spatial coordinates then one typically uses Roman letters for the indices, e.g. $g_{ij}$ where $i,j=1,2,3$.

In Einstein's theory of General Relativity, the effect of gravity is to distort the curvature of space in the presence of matter. The relationship between the curvature and matter is given by {\bf Einstein's equation}
\bea
G_{\mu\nu}\equiv R_{\mu\nu}-\frac{1}{2}g_{\mu\nu}R = 8\pi G T_{\mu\nu}.
\eea
The left hand side of the equation describes the curvature of space: $G_{\mu\nu}$ is the Einstein tensor, $R_{\mu\nu}$  is the Ricci tensor  and $R=g^{\mu\nu}R_{\mu\nu}$ is the Ricci scalar, both of which are functions of the metric $g_{\mu\nu}$ and its derivatives with respect to the space-time coordinates. The RHS is written in terms of $T_{\mu\nu}$, the {\bf energy-momentum tensor}, describing the properties of matter in the universe.

For a perfect fluid, with energy density $\rho$, and isotropic pressure, $P$,
\bea
T^{\mu\nu}=(\rho+P)U^\mu U^\nu+Pg^{\mu\nu}
\eea
where $U^\mu$ is the four velocity of the fluid. In the fluid's rest frame $U^\mu$=diag(1,0,0,0) and 
\bea
T_{\mu\nu}&=& \mathrm{diag}\left(\rho,\frac{P}{a^2},\frac{P}{a^2},\frac{P}{a^2}\right)
\\
{T^{\mu}}_{\nu}&=&\mathrm{diag}\left(-\rho,P,P,P\right) \label{Tmunu}
\eea

The $\mu=\nu=0$ component of Einstein's equation gives us the {\bf Friedmann equation}
\bea
\left(\frac{\dot{a}}{a}\right)^2 &=&H^2 = \frac{8\pi G}{3}\rho-\frac{K}{a^2}.\label{Friedmann}
\eea
where dots denote derivatives with respect to $t$.

The Friedmann equation tells us that a universe containing matter has to be dynamically evolving, either expanding or contracting, it cannot be static.
Commonly one can also find Newton's constant replaced by the reduced Planck mass $M_{p}^2\equiv 1/8\pi G$ (in units where $\hbar=c=1$).

 The combination of the $\mu=\nu=j$ and $\mu=\nu=0$  equations gives us the {\bf acceleration equation}
\bea
\frac{\ddot{a}}{a} &=& -\frac{4\pi G}{3} \left(\rho+3P\right). \label{accel}
\eea
For ``normal" matter the energy density and pressure are both positive for which (\ref{accel}) predicts decelerating cosmic expansion. 
  
{\bf Energy-momentum conservation} allows us to see how the energy density of matter evolves as the universe expands,
\bea
{T^{\mu}}_{\nu;\mu}&=& {T^{\mu}}_{\nu,\mu}+{\Gamma^{\mu}}_{\alpha\mu}{T^{\alpha}}_{\mu}- {\Gamma^{\alpha}}_{\nu\mu}{T^{\mu}}_{\alpha}=0.
\eea
Here $X_{,\mu}\equiv \partial X/\partial x^{\mu}$, a partial derivative with respect to the coordinate $x^\mu$,  $X_{;\mu}\equiv \nabla_\mu X$ denotes the covariant derivative, that preserves parallel transport in a curved or expanding space and $\Gamma^{\alpha}_{\beta\gamma}$ is the Christoffel symbol relating the two derivatives. In flat non-expanding space, all $\Gamma^{\alpha}_{\beta\gamma}=0$, and the two derivatives are identical. The covariant derivative is such that $g^{\mu\nu}_{;\mu}$=0.

 For $\nu=0$ we get the {\bf fluid or ``continuity" equation},
\bea
\dot\rho+3H(\rho+P) =0\label{fluid}
\eea

Two out of these three equations (the Friedmann, acceleration and fluid equations) are independent, the third being able to be obtained from the other two.

\subsection{The Cosmological Constant}\label{Lambda}

In classical, non gravitational physics, the normalization of energy is arbitrary, since the dynamics of particle  moving under some potential, $V(x)$, is the same as that moving through, $V(x)+V_0$. In GR however the actual value of the energy density matters, i.e. the value of $V_0$ has a role in the dynamics. This opens up the freedom to include a {\bf cosmological constant, $\Lambda$}, a component with a constant energy density, into the dynamical equations. Here we discuss the effect of including $\Lambda$ to the dynamical equations, and reserve discussions of its theoretical origins to section \ref{Lamdbatheory}.

If the effect of adding in a constant energy density is to be insensitive to the choice of coordinates then it has to satisfy
\bea
{T^{\mu(\Lambda)}}_{\nu;\mu}=0
\eea
from our discussion in the previous section about the effect of the covariant derivative on the metric, this implies that ${T^{\mu}}_{\nu}\propto {g^{\mu}}_{\nu}$. This means that from (\ref{Tmunu}), ${T^{\mu}}_{\nu}=-\rho_\Lambda {g^{\mu}}_{\nu}$ and $P_{\Lambda}=-\rho_{\Lambda}$. Note $P_\Lambda=-\rho_\Lambda$ also falls out of the fluid equation (\ref{fluid}), requiring $\dot\rho_\Lambda=0$

Including $\Lambda$, as an interpretation as a new type of matter, modifies the RHS Einstein equation
\bea
G_{\mu\nu} = 8\pi G (T_{\mu\nu} -\rho_{\Lambda}g_{\mu\nu})
\eea

Including $\Lambda$ does not destroy the general covariance (coordinate independence) of Einstein's equations. For this reason Einstein was able to include $\Lambda$ to the LHS of his equations,  interpreting it as a mathematical degree of freedom that can be added,
\bea
G_{\mu\nu} +\Lambda g_{\mu\nu}= 8\pi G T_{\mu\nu}.
\eea
Both of these are mathematically and dynamically equivalent.

Adding in a cosmological modifies the Friedmann and acceleration equations
\bea
\left(\frac{\dot{a}}{a}\right)^2 &=&H^2 = \frac{8\pi G}{3}\rho-\frac{K}{a^2}+\frac{\Lambda}{3}
\\
\frac{\ddot{a}}{a} &=& -\frac{4\pi G}{3} \left(\rho+3P\right)+\frac{\Lambda}{3}
\eea
This allows the universe's expansion to be accelerating if $\Lambda$ dominates.

\subsection{The equation of state parameter, $w$}\label{eos}

For a matter species $i$ we can describe the relationship between its pressure and energy density using an {\bf equation of state parameter, $w_i$,}
\bea
w_i \equiv \frac{P_i}{\rho_i}.
\eea
We can write the fluid equation, (\ref{fluid}), in terms of this
\bea
\dot\rho+3H(1+w)\rho =0
\eea
which when integrated yields
\bea
\frac{\rho_i(a)}{\rho_i^0} &=& \exp\left[-3\int_a^1(1+w_i(a))d\ln a\right].
\eea
where $\rho_i^0$ is the energy density today, when $a=1$.

For normal  matter, the pressure of a species is related to the kinetic energy $\propto m<v>^2$, while the energy density $\sim mc^2$, hence for non-relativistic matter, such as baryons and cold dark matter (CDM), one would expect the equation of state to be negligible, while for relativistic species, like photons, $<v^2>\sim c^2/3$.

We can describe these, and $\Lambda$, very well using a constant equation of state
\bea
w_{non-rel} &\approx& 0 
\\
w_{rel} &=& \frac{1}{3}
\\
w_{\Lambda}&=&-1.
\eea
For a constant $w$ (\ref{fluid}) yields 
\bea
\frac{\rho_i(a)}{\rho_i^0} &=& a^{-3(1+w_i)}
\eea

The universe is made up of a variety of different matter species, and to consider the dynamics from their combined effect we can define an effective equation of state, 
\bea 
w_{tot} =\frac{P_{tot}}{\rho_{tot}}
\eea 
which from (\ref{accel}) will give acceleration if $w_{tot}\leq 1/3$.
\subsection{Dark energy}
The cosmological constant can generate acceleration with $w=-1$, however, as we will discuss in detail in subsequent lectures, it is not alone in this. The family of theoretical types of matter that can give rise to acceleration are given the collective name {\bf dark energy}. They are typically described by an equation of state parameter $w$ that is either constant, or time evolving, but that is sufficiently negative to give $w_{tot}<-1/3$  at late times.

\subsection{The critical density, $\rho_{crit}$, and fractional energy density, $\Omega_i$} \label{rhocrit}
The  {\bf critical density, $\rho_{crit}(a)$},  is the total energy density of matter required to give the Hubble parameter $H(a)$ in a flat universe, i.e. $K=0$ in (\ref{Friedmann}),
\bea
\rho_{crit}(a) \equiv \frac{3H^2(a)}{8\pi G}
\eea
The {\bf fractional energy density} in species $i$, $\Omega_i(a)$,is the fraction of the critical density in that species
\bea
\Omega_i(a) \equiv \frac{\rho_i(a)}{\rho_{crit}(a)}.
\eea
The fractional energy {\it today} for a species will be written $\Omega_i^0$.
\subsection{The deceleration parameter, $q$}\label{decelpar}

To complement the Hubble parameter, $H(a)=\dot{a}/a$ we can define an {\bf acceleration parameter, $q(a)$}, to describe the second order change in the expansion history,
\bea
q(a)\equiv -\frac{\ddot{a}}{aH^2}. \label{decelpardef}
\eea
If GR is assumed to hold then the Friedmann and acceleration equations give
\bea
q(a) = \frac{1}{2}\sum_i\Omega_i(a)[1+3w_i(a)], \label{decelpardef2}
\eea
\subsection{Redshift, $z$}
 The {\bf redshift, $z$,} measures the stretch in the wavelength of light due to the expansion of space.  Consider two photons emitted at time $t_e$ and $t_e+dt_e$, and received at $t_r$ and $t_r+dt_er$. Since they have traversed the same comoving distance
\bea
\int_{t_e}^{t_r}\frac{dt'}{a(t')} &=& \int_{t_e+dt_e}^{t_r+dt_r}\frac{dt'}{a(t')} 
\\
\int_{t_e}^{t_e+dt_e}\frac{dt'}{a(t')} &=& \int_{t_r}^{t_r+dt_r}\frac{dt'}{a(t')} 
\\
\frac{dt_e}{a(t_e)}&=&\frac{dt_r}{a(t_r)}
\\
\frac{\lambda(t_r)}{\lambda(t_e)}&=&\frac{a(t_r)} {a(t_e)}= \frac{1}{a}
\eea
Define the redshift
\bea
1+z\equiv \frac{\lambda(t_r)}{\lambda(t_e)} = \frac{1}{a}
\eea
Note this cosmological redshift is caused by the expansion of space, not due to the motion of the galaxies that would exist in the absence of expansion (this is called {\bf peculiar motion}).
\bea
(1+z_{tot}) = (1+z_{cosmo})(1+z_{pec})
\eea
Typical peculiar velocities are of order a few 100 km/s so that, given $H_0$, peculiar velocities are $\sim10\%$ of cosmological expansion rate at distances $\sim$ few tens of Mpc.

\subsection{Comoving distance, $\chi$}\label{dcom}

The {\bf comoving distance, $\chi$} is the distance between two objects {\it instantaneously today}. It is impossible to actually measure this, because of finite speed of light, but nonetheless it is a useful conceptual distance to calculate since other, observable, distances can be simply related to it. 

Consider a light ray purely in the radial direction $d\chi=dr$, $d\theta = d\phi=0$, then
\bea
\chi = \int_0^r dr
\eea
We can relate the distance to the path of a photon, and hence time, using the metric (\ref{frw}). For a photon, $ds^2=0$. If it is emitted by the object at time $t$ and observed today, at time $t_0$,
\bea
a(t) d\chi &=& c dt
\\
\chi(t) &=& \int_{t}^{t_0}\frac{dt'}{a(t')} 
\eea
We can alternatively write this in terms of Hubble factor
\bea
\chi(a) &=& \int_{a}^{1}\frac{da'}{a'^2H(a')} 
\eea
The comoving distance can  then be calculated for a given cosmology using the Friedmann equation
\bea
\chi(a) &=& \int_{a}^{1}\frac{da'}{\sqrt{\sum_i \Omega_i^0 a^{1-3w_i}}}\label{chifromomega}
\eea

The {\bf comoving horizon} is the maximum distance, on the comoving grid that information can have been carried. It is the comoving distance light could have travelled since the start of the universe.
\bea
\tau &=&  \int_{0}^{1}\frac{da'}{a'^2H(a')}
\eea

\subsection{Angular diameter distance, $d_A$}\label{dang}

The {\bf angular diameter distance, $d_A$} is the distance to an object inferred from comparing the angular size of an object to its (assumed known) length. 

We define the angular diameter distance using the naive Euclidean (flat), non-expanding geometry. If the object of length $L$ subtends an angle $\theta$,
\bea
d_A \equiv \frac{L}{\alpha}\label{dAdef}
\eea
We have to take into account that space has expanded between emission and observation. If light was emited at scale factor $a(t)$, then the comoving size today is $l/a$. The angle, $\alpha$, is unaffected by expansion. The comoving distance to the object is, $\chi(a)$, in comoving coordinates (i.e. length scales today) is then
\bea
\alpha &=& \frac{L}{a\chi(a)} \label{angledef}
\eea
hence
\bea
d_A &=& a\chi(a) = \frac{\chi(z)}{1+z}\label{dAofchi}
\eea
If we live in a curved space then the angular size of the object is modified by the curvature, through its effect on the path of the light rays from the object. Consider the object length $L$ spans coordinates $(r,\theta,\phi)$ and $(r,\theta+\delta\theta,\phi)$, then the physical length of the object in a generally curved space, from the RW metric is:
\bea
L &=& a(t) S_{\kappa}(r,R_0)\delta\theta
\eea
In a curved space therefore
\bea
d_A &=& aS_{\kappa}(\chi,R_0) 
\eea

\subsection{Luminosity distance, $d_L$}\label{dlum}
The {\bf luminosity distance, $d_L$} is inferred from comparing the inferred brightness  (or ``flux" measured in  power per unit area) of an object to its (assumed known) luminosity (in units of power). 

 We define the luminosity distance  again using a (naive) Euclidean, non-expanding geometry. If the object of luminosity $L$ is observed to have flux $F$, then using the inverse square law,
\bea
d_L \equiv \sqrt{\frac{L}{4 \pi F }}\label{dLdef}
\eea

Cosmic expansion has two primary effects on the luminosity distance 1) the light is redshifted, the wavelength increases  $\sim1/a$ and therefore the energy decreases by factor $a$, 2) the number of photons arriving at the detector is also reduced by a factor $a$ because of the stretching of space, therefore $L_{obs}=a^2L_{emit}$.  The comoving distance to the object is again $\chi(a)$, hence
\bea
F = \frac{a^2L}{4\pi \chi(a)^2}
\eea
\bea
d_{L} = \frac{\chi(a)}{a} = (1+z)\chi(z) \label{dLfromchi}
\eea

Consider photons emitted from a surface area A. Then if the universe's geometry is curved the effective observed area is $A S^2_{\kappa}(r,R_0)$ i.e. photons are bent outwards, increasing  the effective area, decreasing the observed flux,  in closed geoemetries and the opposite for open geometries:
\bea
d_L &=& \frac{S_{\kappa}(\chi,R_0)}{a}
\eea

A key difference from our naive Euclidean perspective is that different methods of measuring distances yield different results in an exanding universe. \bea
d_{L}(z) = (1+z)^2 d_A(z)
\eea
While this effect becomes negligible at low redshifts it is important at cosmologically relevant scales. 

\section{Evidence for cosmic acceleration from distance measurements}\label{distance}

\subsection{Supernovae}\label{sn1a}

{\bf Type 1a supernovae (SNIa)} gave the first evidence for cosmic acceleration, and have arguably provided the most effective constraints on the dark energy equation of state parameter.  SNIa are believed to be carbon-oxygen white dwarf stars in a binary system that  are accreting matter from a companion. When the white dwarf's mass approaches the Chandrasekhar mass, $\sim 1.4$ solar masses, it undergoes a thermonuclear explosion. 

An object is a {\bf standard candle} if its luminosity  can be inferred from either an understanding of its physical properties, or some other observable. Astronomers typically refer to {\bf magnitudes} rather than fluxes and luminosities when presenting observational data. An object's {\bf apparent magnitude, $m$, } is related logarithmically to its  flux. For two objects, their apparent magnitudes and fluxes are related by
\bea
m_1-m_2 &=& 2.5\lg \frac{F_2}{F_1}
\eea
with the zero-point $m=0$ set by a reference star, Vega. The Sun's apparent magnitude is $m_{Sun}=-26.7$. 
An object's {\bf absolute magnitude} is related to the absolute luminosity of the object. It is defined as its apparent magnitude if it were a fixed distance, 10pc, away from us. Using the definition of the luminosity distance
\bea
m-M &=& 5\lg \frac{d_L}{10pc}.
\eea
The Sun's absolute magnitude is 4.83.

 The {\bf lightcurve} for a SNIa explosion (the apparent magnitude $m$, versus time) is predominantly determined by radioactive decays of two species. The decay of $^{56}Ni$ dominates  in the first few days/weeks, and determines the peak luminosity of the event.  The lightcurve decay, in the latter weeks, is determined by the decay of $^{56}Co$.  If the white dwarf combustion is complete then it is expected that roughly $\sim 0.6$ solar masses of $^{56}Ni$ is produced. As such, one might expect all SNIa events to have comparable luminosities  and for them to be strong {\bf standard candle} candidates. Since the typical peak luminosity of a SNIa is a few billion times that of our Sun ($M_{SN}\approx -19.3$,  implying $L_{SN}/L_{sun}=10^{(M_{sun}-M_{SN})/5}\sim 10^{9}-10^{10}$), these events can be as bright as their host galaxy, providing an effective distance measure to large cosmic distances.  In reality there is an intrinsic variation in the peak luminosities of SNIa which could threaten their usefulness as a standard candle. However an empirical relation between the SNIa peak brightness and the rate at which the lightcurve decays after the peak has been found, the {\bf Phillips relation}\cite{Phillips:1993}, in which intrinsically brighter supernovae decay more slowly. This relation allows the luminosity to be inferred from fitting the lightcurve, reducing the intrinsic spead.

From (\ref{dLdef}), the comparison of the observed apparent magnitude, $m$, to the presumed absolute magnitude yields an estimate of $d_{L}$. Consider a Taylor expansion of the expansion rate close to today (time $t=t_0$),
\bea
a(t) = 1+H_0(t-t_0)-\frac{q_0H_0^2}{2}(t-t_0)^2+...
\eea
where $q_0$ is the value of the deceleration parameter (\ref{decelpardef}) today. By writing $a(t)$ in terms of $(1+z)^{-1}$ and inverting the expression to find $t(z)$, one can obtain a Taylor expansion of $d_L$  (\ref{dLfromchi}),
\bea
H_0d_L(z) = z+\frac{1}{2}(1-q_0)z^2+...
\eea
From (\ref{decelpardef2}), $q_0=\Omega_m/2-\Omega_\Lambda$ for a $\Lambda$CDM universe and  a measurement of $q_0$ would lead to a degeneracy  in the $\{\Omega_m, \Omega_\Lambda\}$ parameter space described by $\Omega_m/2-\Omega_\Lambda=$ constant. One wouldn't be able to constrain $\Omega_m$ and $\Omega_\Lambda$ separately. Of course the Taylor expansion breaks down at higher $z$ so that the constraint is not exactly of the form we see here, however a degeneracy still persists, and one needs other, complementary observations to break it.

As an example, let's consider 2 supernovae:
\bea
1992P: & z=0.026, & m=16.08
\\
1997ap: & z=0.83, & m=24.32.
\eea
For $z\ll 1$ then $d_L(z)\approx z/H_0$, and assuming a value for $H_0=72km/s/Mpc^{-1}$, one finds the absolute magnitude of the low $z$ SNIa to be $M=-19.09$. Assuming this also holds for the higher supernova gives $H_0d_L(z=0.83)\approx 1.16$. One can then compare this to the theoretical estimate for $H_0d_L$ for a pure matter universe $\Omega_m=1$, for which $H_0d_L(z=0.83)\approx0.95$, and for a $\Lambda$CDM universe, $\Omega_m=0.3, \Omega_\Lambda=0.7$, for which $H_0d_L(z=0.83)\approx1.23$. 

Observational results for luminosity distances are often given using the {\bf distance modulus, $\mu$}, 
\bea
\mu &\equiv& m-{\mathcal M} = 5 \lg \left(H_0 d_L\right) 
\eea
with
\bea
{\mathcal M} &\equiv& M-5\lg (H_0 Mpc^{-1})+25.
\eea
where ${\mathcal M}$ is calibrated using low-redshift supernovae.

In reality, the measured apparent magnitude can be affected by numerous systematic errors including extinction and redenning from dust in the galactic host.  The magnitude is measured over a finite frequency range,  or passband. The distribution of the magnitude over the different passbands is referred to as the supernovae's  color. Redenning can cause photons to be shifted from one passband to another which can lead to over or under-estimation of the apparent magnitude if a limited number of passbands are used. If the intrinsic colors of the SNIa or the extinction law were precisely known, then the extinction can be eliminated from the distance modulus by measurements of multiple passbands. Realistically however, intrinsic variations in galactic dust properties and SN colors twinned with errors in the measurements of the apparent magnitude (photometric errors) will still lead to uncertainties in the distance modulus. A way to mitigate this is to measure the supernovae in the near-infrared, where the effects from dust extinction are not as significant.

The accuracy of the distance modulus is also dependent on the low-redshift sample of supernovae used to ``anchor" the Hubble diagram, and calculate ${\mathcal M}$. Until recently this sample has been rather small and heterogenous and has correlated large-scale, correlated peculiar velocities. The peculiar velocities obscure extraction of the cosmological velocity and lead to errors in the distance estimates. Currently a number of surveys are underway to increase the low redshift sample,  including the Harvard Center for Astrophysics supernovae project \cite{Hicken:2009df}, the Carnegie Supernova project, the Nearby Supernova Factory and the Sloan Digital Sky Survey II supernova survey.

As supernovae at higher and higher redshift are observed the issue of evolution in supernovae luminosities arises. There is an observed correlation between peak brightness and host-galaxy type, and with host-galaxy properties (such as star formation rate and metalicity, and redshift). This evolution sensitivity can be reduced by using Phillips-corrected lightcurves are less sensitive to galaxy-host environment and finding low-redshift analogs of higher-redshift galactic environments  to calibrate supernovae behavior. Ultimately though a better understanding of supernovae physics will be required to reduce sensitivity to supernovae evolution and the effect of environment.

\subsection{Cosmic Microwave Background (CMB)}\label{cmb}

The Cosmic Microwave Background (CMB) are relic photons released at the time of {\bf recombination}, when the ionized plasma of free electrons and protons present in the early universe cools enough for neutral hydrogen to form. While photons readily scatter of free electrons through Thomson scattering they interact  negligibly with neutral hydrogen, so post-recombination the photons free-stream, with minimal interaction until they are detected today.  

The CMB photons are remarkably homogeneous in temperature across the whole sky. However small fluctuations in the temperature are present at the level of $10^{-5}$, or a few $\mu K$ in magnitude. The utility of the CMB in placing constraints on dark energy comes from considering the characteristic scale over which these fluctuations should be correlated.

The CMB provides a complementary measure of the expansion history to the superanovae. While supernovae are standard candles, the CMB is said to be a {\bf standard ruler}.  The  angular diameter distance is inferred by comparing the apparent angular size of correlations in the CMB temperature to their expected intrinsic size, estimated from understanding the propogation speed of photons in the ionized plasma, prior to recombination.

Recombination takes place when the photons temperature is $T\sim 3000K$. As the universe expands the photons redshift and cool, and have a measured temperature today of $T=2.726K$. This implies that recombination occured when the universe was  roughly 1100 times smaller than it is today, $z_{rec}\sim 1100$.  For the standard cosmological composition, this was when the universe was $\sim 400,000$ years old.

The speed of photons in the plasma prior to recombination, the sound speed $c_s$, is roughly given by  $c_s=c/\sqrt{3}$. At recombation the sound horizon, $r_s$, describes the maximum distance photons could have travelled, and hence a typical correlation length for any inhomogeneities (temperature fluctuations) that could be generated through causal processes. The  comoving scale of the sound horizon is given by 
\bea
r_s = \int c_s d\tau \sim \frac{(1+z_{rec})c_s}{H(z_{rec})}
\eea
which is $\approx145 Mpc$ for a standard cosmological composition. Since recombination occurs early in the universe's history one finds   that curvature and, in most scenarios, dark energy have negligible effects on the sound horizon. The apparent angular size of the correlations, $\theta$, is then determined to the angular diamater distance to recombination, following (\ref{dAdef}),
\bea
\theta &=& \frac{r_s}{d_A}
\eea
 The universe's curvature/ geometry  could have a significant effect on the angular size these correlations subtend, with an open geometry decreasing $\theta$, and a closed geometry increasing it. The main constraint on the $\{\Omega_m,\Omega_\Lambda\}$ parameter space coming from the CMB is therefore a constraint on $\Omega_{tot} = \Omega_m+\Omega_\Lambda$, complementary to the measured combination from the supernovae. Since late time acceleration increases the angular diameter distance, adding dark energy or $\Lambda$ has a geometric degeneracy with curvature. The degeneracy can be broken however by including an additional constraint on the value of $H_0$.

The CMB contains more information than just the angular diameter distance that is sensitive to the properties of dark energy. This is from the  Integrated Sachs Wolfe (ISW) effect related to the growth of large scale structure. We will discuss this later in section \ref{isw}.

\subsection{Baryon Acoustic Oscillations (BAO)} \label{bao}

\begin{figure}
\begin{center}
\includegraphics[width=5in,angle=0]{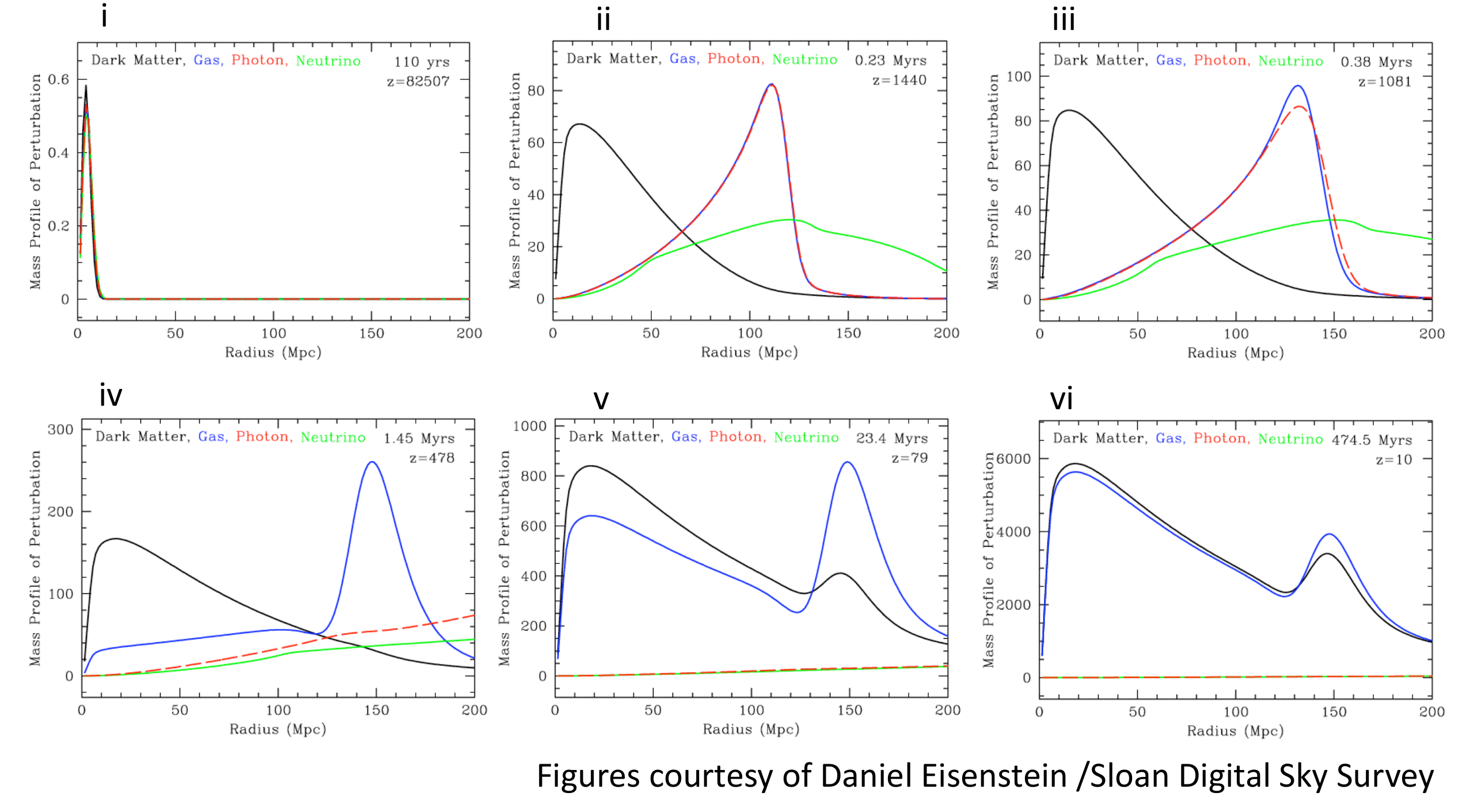}
\caption{Figures i-vi show the evolution of a delta-function spike in the baryon (labeled ``gas") density over time along with analogous evolution for dark matter, photon and neutrinos. Figure i: An initial delta-function spike, early in the universe's history, well before recombination occurs.  Figures ii and iii: the photon and gas over-densities are tightly coupled up to recombination, with both propogating at speed $c_s$. At decoupling the correlation length (here depicted by the comoving radius of the overdensity ``ripple") is $\sim 150Mpc$. Figure iv: following decoupling, the acoustic oscillation correlations in the baryon density distribution do not propogate further and the radius of the correlation remains fixed while the photons continue to propogate at the speed of light. Figure v: the dark matter and baryons now fall into the other's gravitational well. Figure vi: today we see correlated density distributions of baryons and dark matter that include the correlation scale imprinted from the baryon photon coupling prior to recombination. [Credit: Daniel Eisenstein and the Sloan Digital Sky Survey collaboration.]}
\label{rb_fig2}
\end{center}
\end{figure}

Prior to recombination the photons and baryons  (mostly protons) were tightly coupled to one another, through Thomson scattering, and the electrostatic attraction of the electrons and protons. Fluctuations in the baryon density were able to propogate at comparable speeds to the photons, i.e. with $c_s\sim c/\sqrt{3}$. When recombination occured the CMB photons  ``decoupled" from matter. Since the baryon mass is much larger than their thermal energy the particles are highly non-relativistic and their sound speed  dropped effectively to zero. After decoupling, a correlation with characteristic size $r_s$ remains imprinted in the distribution of baryons. Figure \ref{rb_fig2} gives a pictorial summary of how the characteristic scale is imprinted in the baryon and dark matter matter distribution we observe today. 

Fluctuations in the baryon density seed the galaxies and clusters of galaxes, {\bf large scale structure}, and the galaxy correlation function that we observe today contains a preferred comoving scale $\sim r_s$. When Fourier transformed, the correlation gives rise ripples in the power spectrum of the baryon density fluctuations. This correlation is therefore known as {\bf baryon acoustic oscillations}.

\begin{figure}
\begin{center}
\includegraphics[width=3.5in,angle=0]{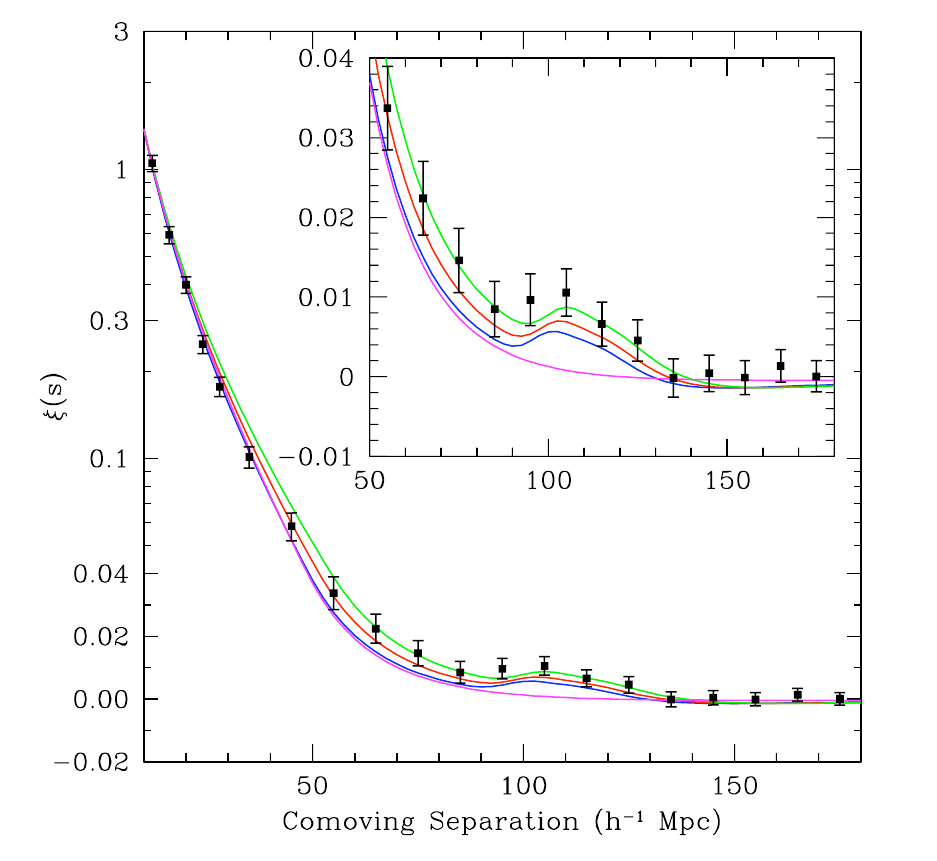}
\caption{The galaxy correlation function, $\xi$, from the Sloan Digital Sky Survey \cite{Eisenstein:2005su} clearly showing the baryon acoustic correlation at 100$h^{-1}Mpc$ ($H_0=100hkms^{-1}Mpc^{-1}$ so that a typical value for $h\sim 0.7$).}
\label{rb_fig3}
\end{center}
\end{figure}

The  auto-correlation in galaxy clustering  is used to extract the correlation from a galaxy survey,
\bea
\xi(s) = \left\langle \frac{\delta\rho}{\rho}(\bf{x_1})\frac{\delta\rho}{\rho}(\bf{x_2})\right \rangle
\eea
where $\rho$ and $\delta\rho$ are the homogeneous and fluctuation in the matter denstiy and $\langle...\rangle$ is the average over all points in the sky such that $|{\bf x_1}-{\bf x_2}|=s.$  Figure \ref{rb_fig3}  shows the correlation function\cite{Eisenstein:2005su}  for 46,748 luminous red galaxies from the Sloan Digital Sky Survey. 

Since galaxy surveys are 3-dimensional, the correlation can be seen in both the radial (line-of-sight) distribution, $r_{\parallel}$, and the angular (transverse) correlations, $r_{\perp}$, of the galaxies. These are effected differently by the universe's expansion
\bea
r_{\parallel}(z)& =& \frac{c}{H(z)}\Delta z
\\
r_{\perp}(z)& =&d_A(z)\Delta \theta.
\eea
The 3D correlations constrain a combination of these
\bea
D_V(z) = r_\perp^2r_\parallel = \frac{c}{H(z)}d_A(z)^2.
\eea
which can then be used to constrain cosmological parameters.

The Sloan Digital Sky Survey \cite{Eisenstein:2005su}  place constraints on $D_V$ at the effective redshift of the galaxies in the sample, $z=0.35$,
\bea
A(z=0.35) \equiv \frac{D_V(z=0.35)\sqrt{\Omega_m^0H_0^2}}{2c= 0.469\pm 0.017}
\eea

The baryon acoustic oscillations, because they have a characteristic size that can be determined from an understanding of the physics of recombination, are an additional, and complementary, standard ruler to the CMB fluctuations. However while the CMB is a standard ruler to $z\approx 1100$, the BAO provide a standard ruler at the redshift of the galaxies used to measure the correlation, at $z\lesssim 1$. As such, they provide information, complementary to that from supernovae, about the evolution of the expansion history at the redshifts when cosmic acceleration is occurring. 

\subsection{Complementarity of the distance measures} \label{complement}

\begin{figure}
\begin{center}
\includegraphics[width=2.5in,angle=0]{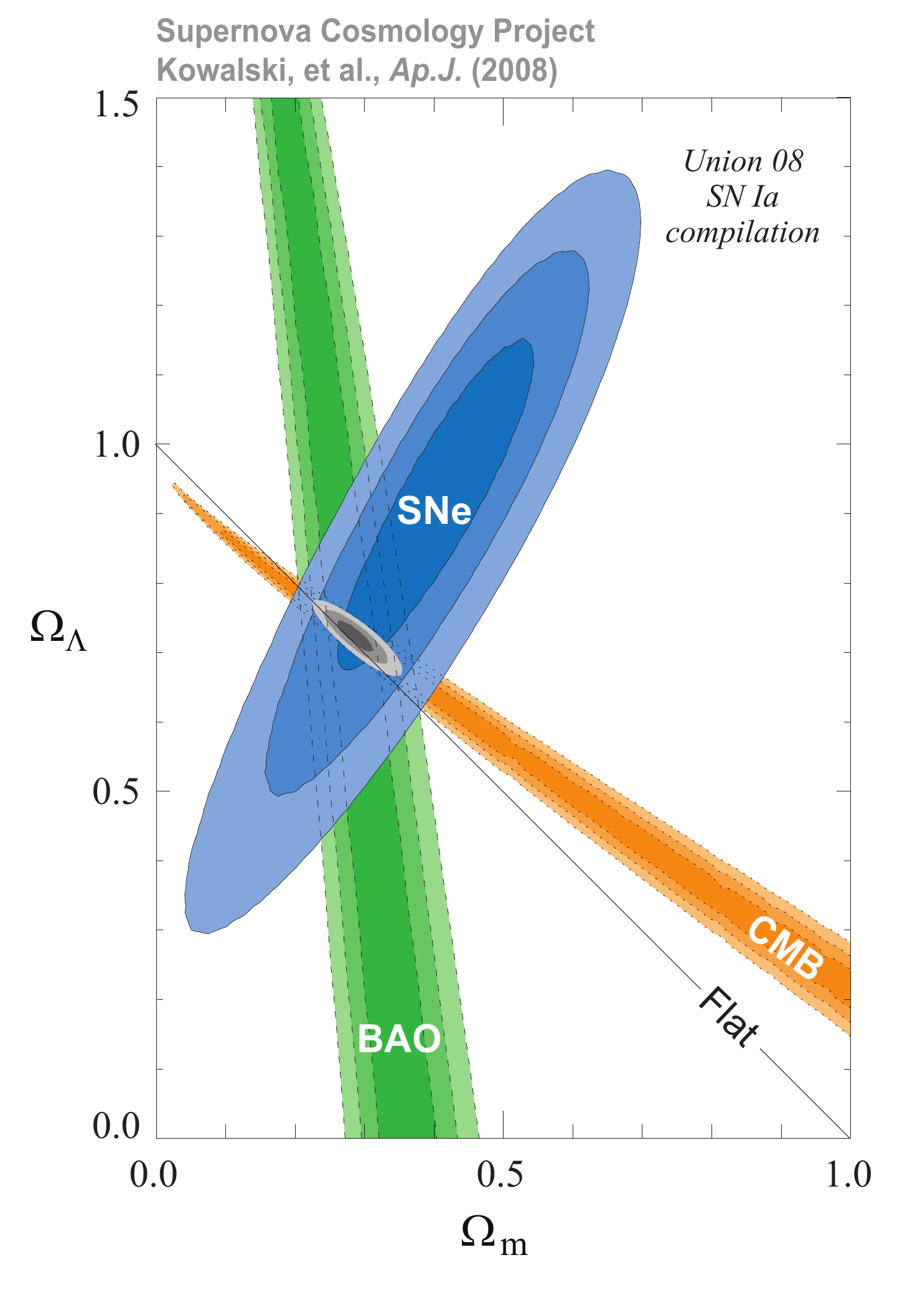}\includegraphics[width=2.5in,angle=0]{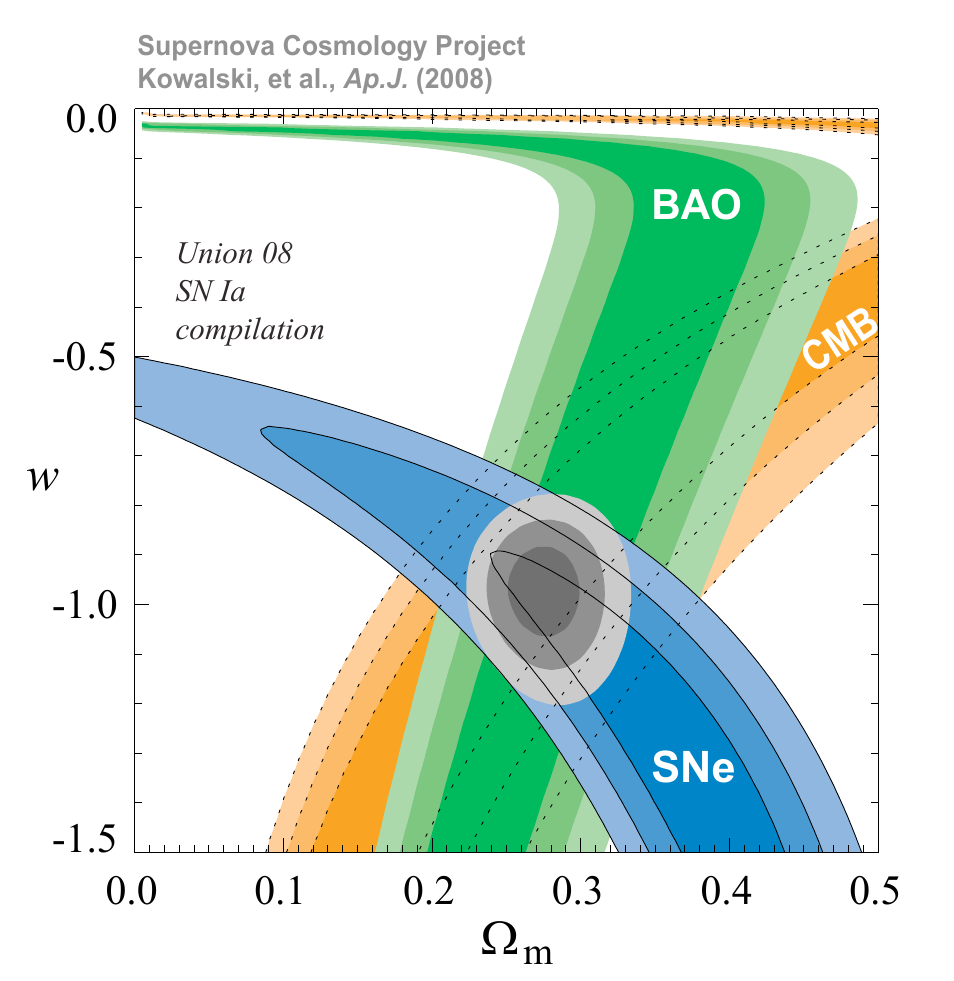}
\caption{Combined constraints in the $\Omega_m-\Omega_\Lambda$ (left panel) and $\Omega_m-w$ (right panel) parameter spaces  for the WMAP 5-year CMB data, SDSS BAO and Supernova Cosmology Project (SCP) Union supernovae constraints \cite{Kowalski:2008ez}.}
\label{rb_fig4}
\end{center}
\end{figure}

In figure \ref{rb_fig4} we show the complementarity of the  supernovae, CMB angular diameter distance and BAO datasets, with recent constraints on the fractional energy densities of matter and $\Lambda$, and the dark energy equation of state\cite{Kowalski:2008ez}.  The three provide consistent, overlapping confidence level contours, that in combination indicate a concordance cosmology with $\Omega_m = 0.279\pm0.015$, $\Omega_\Lambda = 0.721\pm0.015$ at 68\% confidence level for a flat $\Lambda$CDM cosmology, and $ -0.097<w+1<0.0142$ at the 95\% confidence level for a flat universe with a constant dark energy equation of state \cite{Komatsu:2008hk}. 

\begin{figure}
\begin{center}
\includegraphics[width=5in,angle=0]{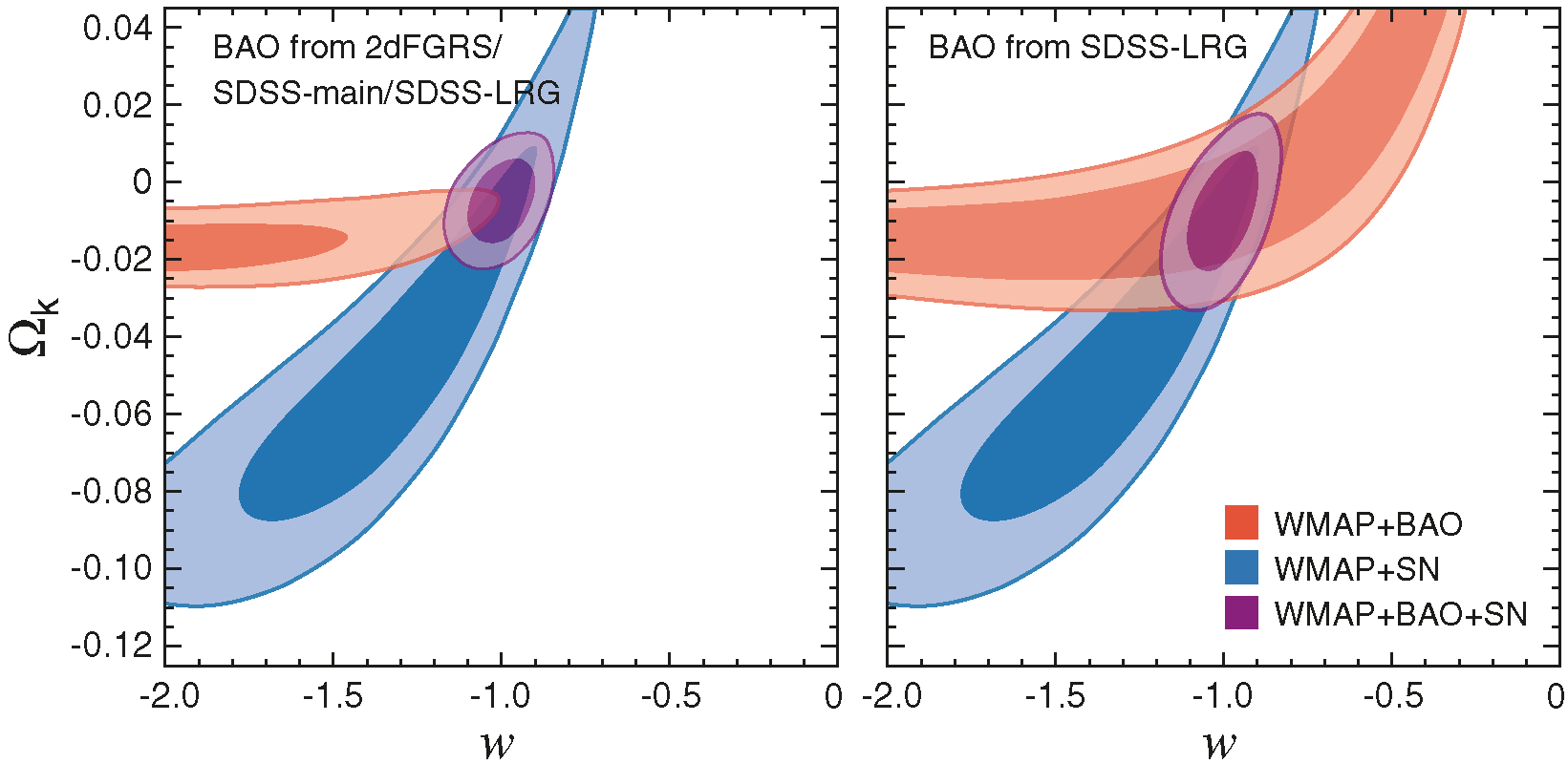}
\caption{Combined constraints in the $\Omega_k-w$ parameter spaces  for the WMAP 5-year CMB data, SDSS and 2 degree field (2dF) BAO measurements and Supernova Cosmology Project (SCP) Union supernovae constraints \cite{Komatsu:2008hk}.}
\label{rb_fig5}
\end{center}
\end{figure}

Figure \ref{rb_fig5} shows the constraints on curvature and the dark energy equation of state if the assumption of flatness is relaxed.

\section{Theoretical avenues to explain cosmic acceleration}\label{theory}

\subsection{The Cosmological Constant revisited}\label{Lamdbatheory}

We have already seen in section \ref{Lambda} that a Cosmological Constant can provide accelerated expansion if it dominates the energy density.

In classical General Relativity, $\Lambda$ is a constant of nature able to be added into the Einstein-Hilbert action for gravity without destroying covariance
\bea
S_{EH} = \int d^4x\sqrt{-g}\frac{1}{16\pi G}(R-2\Lambda).
\eea

Einstein introduced $\Lambda$ to create a static universe (at the time astronomical observations had not yet extended out of the galaxy, and so there was no evidence for cosmic expansion). Considering the expansion and acceleration rates today  and requiring both to be zero,
\bea 
\frac{8\pi G}{3}\rho + \frac{\Lambda}{3}-\frac{c^2\kappa}{R_0}^2 &=& H^2 = 0
\\
\Lambda - \frac{4\pi G}{3}\rho &=&\frac{\ddot{a}}{a} = 0
\eea
 gives a constraint relating the curvature and value of $\Lambda$, $R_0=c/\sqrt{\Lambda}$ with $\kappa=+1$ i.e. a closed universe.

Two quotes from Einstein convey his feelings about the introduction of $\Lambda$
\begin{quote}
{\it ``I have again perpetrated something related to the theory of gravitation that might endanger me of being committed to  a madhouse."}
\end{quote}

\begin{quote}
{\it ``[$\Lambda$ is] gravely detrimental to the beauty of the theory."}
\end{quote}

From a classical perspective, as a constant of nature, it is somewhat philosophical to ask why $\Lambda$ has the value it does. However from a quantum perspective the value has great meaning and its observed value causes a major puzzle.

If we consider $\Lambda$ originating in the vacuum energy, then one can use back-of-the-envelope arguments to estimate its expected size. The free-field vacuum energy for a mode of frequency $\omega$ is $\hbar\omega/2$ where $\omega=\sqrt{k^2+m^2}$ with $k$ the wavenumber of the mode and $m$ the mass of the particle. The total vacuum energy is then the sum over the vacuum energy over all modes and all particles
\bea
\rho_{vac} &=&\frac{1}{2}\sum_{particles}g_i \int_0^{k_{max}}\frac{d^3k}{(2\pi)^3}\sqrt{k^2+m^2} \sim\sum_i \frac{g_i k_{max}^4}{16\pi^2}
\eea
where $g_i=(-1)^{2j}(2j+1)$ is the degeneracy factor for a particle of spin $j$, with $g_i>0$ for bosons and $g_i<0$ for fermions. $\rho_{vac}$ is therefore quartically divergent in $k_{max}$. The question now arises, what is $k_{max}$?

\begin{itemize}
\item Pauli  in the 1930s calculated the effect of a photon zero point energy on the curvature of space and already saw problems. If $\lambda_{max}$ is the classical radius of an  electron $\sim 10^{-15}m$ then the contribution to $\Lambda=\rho_{vac}^{\gamma}=\omega_{max}^4/8\pi^2$ implies $R_0\sim 31km$ and {\it `` one could not even reach the moon"}!

\item If $R_0\sim H_0^{-1}$ so that $\rho_{vac}\sim \rho_{crit}\sim 8\times 10^{-10}Jm^{-3}\sim 10^{-47}GeV^4$, as we observe, then that implies that $k_{max}\sim10^{-2}eV$ (or a cut off scale $\lambda_{min}\sim 10^{-4}m$)  well below scales at which we believe we fully  understand physics.

\item If the cut-off is at the Planck scale, then 
\bea\Lambda c^2 \sim  \frac{M_{pl}c^2}{L_{pl}^3}\sim \frac{2\times 10^{9}J}{(1.6\times 10^{-35}m)^3}\sim 10^{113}Jm^{-3}\sim  10^{74}GeV\sim 10^{123}\rho_{crit}\hspace{0.5cm}
\eea
 If one is more conservative about the upper energy limit  assuming a cut-off at the QCD scale, $\Lambda\sim 10^{-3}GeV^4\sim 10^{44}\rho_{crit}$. Invoking supersymmetry provides a way to generate zero vacuum energy while SUSY is unbroken, since bosonic or fermionic superpartners exist for each Standard  Model particle, with opposite degeneracy factor, so that $\rho_{vac}=\rho_{boson}-|\rho_{fermion}| =0$. When SUSY is broken at $T_{SUSY}\sim 1TeV$ then the  contribution to the vacuum energy would be $\Lambda\sim \rho_{vac}^{SUSY}\sim 10^{12}GeV^4\sim 10^{60}\rho_{crit}$.
 
All the options above lead to a phenomenal discrepancy between the theoretical quantum value of $\Lambda$ and the observed value. It is highly unlikely that a classical contribution to $\Lambda$ would exactly this quantum contribution precisely enough to reconcile theoretical and observed values-- this is known as the {\bf $\Lambda$ fine-tuning problem}.
\end{itemize}

Another puzzle associated with the observed value of $\Lambda$ is the {\bf Coincidence problem} -- why are the densities of matter and $\Lambda$ comparable today? We see that observational constraints suggest that matter comprises 30\% of the energy budget today while $\Lambda$ makes up most of the remainder. Why are they so similar given that they evolve so very differently ($\Lambda$ is undiluted by the universe's expansion while matter is diluted as $a^{-3}$)? 

If $\Lambda$ were maybe as little as 10 times larger than the observed value today then it would have catastrophic implications for the development of life, because the earlier onset of acceleration would suppress or prevent galaxies  formation, $\Lambda<0$ would also have poor consequences since the universe  would quickly recollapse. In light of this, some authors have argued  the value of $\Lambda$ is influenced by the {\bf anthropic principle}. Here the value of $\Lambda$ can vary from one region of space to another, and from one universe to another, and the value in our univesre reflects the fact that our universe was able to create galaxies and life \cite{Linde:1986,Weinberg:1987,Vilenkin:1996,Martel:1998,Garriga:1999bf,Garriga:2002tq} . Associated with the anthropic arguments are recent developments in the study of string theory vacua. It appears that there could be a very large number of possible vacua ($>10^{100}$) that can arise from the role of gauge field and brane configurations in the compactification of extra dimensions\cite{Dasgupta:1999ss,Bousso:2000xa,Feng:2000if,Giddings:2001yu,Kachru:2003aw,Susskind:2003kw} .

Raphael Bousso's TASI lectures from 2007 are a good resource \cite{Bousso:2007gp} for a more in depth discussion of the theoretical origins of the cosmological constant .

\subsection{Scalar field dark energy} \label{scalar}

In the absence of a robust resolution to the fine-tuning and coincidence problems, alternatives to $\Lambda$ have been considered  to give rise to cosmic acceleration. One of the most widely investigated ideas is that a new type of matter, a scalar field known as {\bf quintessence}, is driving acceleration. 

The properties of the scalar field can be succintly described by an action, $S$,
\bea
S &=& \int d^4 x \sqrt{-g} {\mathcal L}, \ \ \ \
{\mathcal L} =\frac{1}{2} g^{\mu\nu}\partial_\mu\phi\partial_\nu\phi - V(\phi)
\eea
where $\partial_\mu\phi \equiv \partial\phi/\partial x^{\mu}$ and $V(\phi)$ is a self-interaction potential driving the dynamics of the scalar field.

The energy momentum tensor for the scalar field can be obtained by varying the action with respect to the metric
\bea
T^{(\phi)}_{\mu\nu}&\equiv& -\frac{2}{\sqrt{-g}}\frac{\partial(\sqrt{-g}{\mathcal L})}{\partial g^{\mu\nu}}
\\
&=&\partial_\mu\phi\partial_\nu\phi-g_{\mu\nu}\left(\frac{1}{2} g^{\alpha\beta}\partial_\alpha\phi\partial_\beta\phi - V(\phi)\right)
\eea
We will assume that the scalar field is homogeneously distributed in space so that $\partial_i\phi=0$, in which case considering the 00 and ii components of $T_{\mu\nu}^{(\phi)}$ give the scalar field's energy density and pressure,
\bea
\rho_{\phi} &=& \frac{1}{2}\dot\phi^2+V
\\
P_{\phi}&=&\frac{1}{2}\dot\phi^2-V
\\
w_{\phi}&=& -1+\frac{\dot\phi^2}{\dot\phi^2+2V}
\eea
Acceleration is achieved if the scalar field is slowly evolving, $\dot\phi^2\ll V$ and dominates the energy density.

The equation of motion for the scalar field can be obtained from the varying the action with respect to the field itself, through the Euler-Lagrange equation,
\bea
\frac{\partial(\sqrt{-g}{\mathcal L})}{\partial \phi}-\partial_\mu\frac{\partial(\sqrt{-g}{\mathcal L})}{\partial\partial_\mu\phi}=0
\eea
 which in the FRW metric gives
 \bea
 \ddot\phi+3H\dot\phi+V'=0
 \eea
 where $V'=dV/d\phi$, and $3H\dot\phi$ is ``Hubble drag", an effective frictional term due to the expansion acting against evolution in $\phi$. At the terminal velocity under the action of this friction term, $\ddot\phi=0$ and $\dot\phi\sim V'/3H$ so that the condition for acceleration is dependent on the form of the potential, $\dot\phi^2/V\sim(V'/V)^2\ll1$.
 
 The behavior is effectively a low-energy analogue of the inflaton in the early universe, with the advantage that we don't have to address the ``graceful exit" problem of how inflation is brought to an end, but with the disadvantage that it is far more difficult to motivate the existence and dynamical evolution of a scalar field in the low-energy universe today.
 
Nevertheless, the aim of introducing quintessence is to use ${\mathcal L}(\phi)$ to address the fine-tuning and coincidence problems. Let us examine the behavior of $\phi$ for a couple of potentials in this regard.

A power law potential, $V=M^{4+\alpha}/\phi^{\alpha}$, gives rise to evolution of the scalar field that follows a {\bf dynamical attractor}, that gives accelerated expansion at late times without  independently of the initial conditions for the scalar field  \cite{Zlatev:1998tr,Steinhardt:1999nw} . For the scalar field energy density to be comparable to the critical density today $M\sim 1GeV$ for $\alpha=2$ compatible with high energy astrophysics. This would in some way mitigate the fine-tuning problem though it still remains to be seen if such potentials are able to be convincingly motivated from high energy particle physics. 

An exponential potential, $V=m^4e^{-\lambda\phi/M_p}$,  gives acclerated expansion in the absence of matter, but tracks the scaling behavior of matter, i.e. $w_\phi=w_{mat}$ otherwise \cite{Ferreira:1997au,Ferreira:1997hj,Copeland:1997et} . This can mitigate the coincidence problem, since the fractional energy density of the scalar is a fixed fraction of the background, dominant matter, $\Omega_\phi=3(1+w_{mat})/\lambda^2$, however there then remains the issue of how to generate acceleration at late times.  Acceleration can be induced through extending the scenario by having a sum of two exponentials, including a feature to the exponential potential (to capture and slow the scalar) \cite{Albrecht:1999rm}, or by adding in a non-minimal coupling between the scalar and matter. The review by Copeland et al. \cite{Copeland:2006wr} provides a comprehensive summary of many possibilities explored.

Because,  for the exponential potential, the scalar energy density  has a fixed fractional energy density  it opens up the possibility that, unlike $\Lambda$, it could make up a significant fraction of the energy density at early times (prior to acceleration starting). This is known as {\bf early dark energy}. One can place upper limits on the contribution of $\phi$ at early times both at Big Bang Nucleosynthesis and at recombation \cite{Bean:2001wt}. At BBN, the scalar behaves like relativistic particles with $w=1/3$, and boosts the expansion rate as an additional relativistic energy density. This alters the expected abundances of deuterium and $^4He$ and compared to observed abundances and one finds the constraint $
\Omega_\phi(1MeV)<0.045$.

\subsection{General scalar field models} \label{generalscalar}

A variety of extensions to the minimally coupled scalar field model discussed in \ref{scalar} have been considered. At its most general we can consider an action of the form
\bea
S &=& \int d^4x\sqrt{-g}\left[\frac{f(\phi)}{16\pi G}{\mathcal R}+{\mathcal L}_\phi(\phi,X)+{\mathcal L}_m(\phi)\right]
\eea
where $X=g^{\mu\nu}\partial_\mu\phi\partial_\nu\phi/2$. The first term represents a non-minimal coupling between $\phi$ and gravity, the second is a non-canonical Lagrangian for the scalar field, and the final term is a non-minimal coupling between normal matter and $\phi$. We will consider each of these possibilities separately.

\subsubsection{Modified gravity} \label{modgrav}
While $\Lambda$ and quintessence address cosmic accleration through a modification to the right hand side of Einstein's equation, an alternative might be that GR is modified on scales $c/H_0^{-1}$.

{\bf f(R) gravity:} This is a class of modified gravity models in which the gravitational action contains a general function $f(R)$ of the Ricci scalar.
\be
S=\frac{M_p^2}{2}\int d^4 x\sqrt{-g}\, \left[R+f(R)\right] + \int d^4 x\sqrt{-g}\, {\cal L}_{\rm m}[\chi_i,g_{\mu\nu}] \ ,
\label{jordanaction}
\ee
Varying the action with respect to the metric gives the Einstein field equations which now contain additional terms on the LHS 
\be \label{jordaneom}
\left(1+f_R \right)R_{\mu\nu} - \frac{1}{2} g_{\mu\nu} \left(R+f\right) + \left(g_{\mu\nu}\Box 
-\nabla_\mu\nabla_\nu\right) f_R =8\pi G T_{\mu\nu} \ ,
\ee
where $f_R\equiv df/dR$.

The Friedmann equation and acceleration equations are modified,
\bea
H^2+\frac{f}{6}+H\dot{f}_R&=&  \frac{8\pi G}{3}\rho
\label{jordanfriedmann}
\\
\frac{\ddot{a}}{a}-H^2f_R+a^2\frac{f}{6}+\frac{3}{2}H\dot{f}_R+\frac{1}{2}\ddot{f}_R&=& -\frac{8\pi G}{6}(\rho+3P) \ .
\eea
The extra terms in the acceleration equation are able to reconcile the observed acceleration $\ddot{a}>0$ with a universe populated by matter with positive pressure.

There are a number of conditions on suitable forms of the function $f(R)$: i) $f_{RR}>0$ so that the dynamical behavior is stable in the high-curvature regime \cite{Sawicki:2007tf}, ii) $1+f_R>0$ for all $R$ so that the effective value of Newton's constant, $G_{eff}=G/(1+f_R)$, is positive (so that gravitons are not ghost-like), and iii) $f_R<0$ so that one recovers GR in the early universe.

The action (\ref{jordanaction}) is known as the {\bf Jordan frame}. There exists a complementary, and sometimes conceptually simpler, way in which to approach $f(R)$ modifications to GR known as the {\bf Einstein frame}.  The two are related by a conformal transformation on the metric, 
\bea
\tilde{g}_{\mu\nu} = (1+f_R)g_{\mu\nu}
\eea
so that in the Einstein frame  the gravitational action in the usual Einstein Hilbert form of GR.  The price one pays for this simplification is a non-minimal coupling between matter fields and the new metric \cite{Amendola:1999er,Bean:2000zm}, as well as the appearance of a new scalar degree of freedom evolving under a potential determined precisely by the original form of the $f(R)$ coupling in the Jordan action. 

The gravitational action~(\ref{jordanaction}) can be recast into a dynamically equivalent form \cite{Chiba:2003ir,Magnano:1993bd} in the Einstein frame
\bea
S&=&\frac{M_p^2}{2}\int d^4 x\sqrt{-{\tilde g}}\, {\tilde R} 
+\int d^4 x\sqrt{-{\tilde g}}\, 
\left[-\frac{1}{2}{\tilde g}^{\mu\nu}(\tilde{\nabla}_{\mu}\phi)\tilde{\nabla}_{\nu}\phi -V(\phi)\right] \nonumber
\\ &&+\int d^4 x\sqrt{-{\tilde g}}\, e^{-2\beta\phi/M_p} {\cal L}_{\rm m}[\chi_i,e^{-\beta\phi/M_p}{\tilde g}_{\mu\nu}]\ ,
\label{einsteinaction}
\eea
where $\beta\phi \equiv M_p\ln(1+f_R)$, with $\beta \equiv\sqrt{2/3}$, and the potential $V(\phi)$ is determined entirely by the original form~(\ref{jordanaction}) of the action and is given by
\be
V(\phi)=\frac{M_p^2}{2}\frac{R f_{R} - f}{(1+f_{R})^{2}} \ .
\label{einsteinpotential}
\ee
The Friedmann and acceleration equations in the Einstein frame take the standard forms in GR:
\bea
\tilde{H}^2 &=& \frac{1}{3M_p^2}\left[\tilde\rho(\phi) +\frac{1}{2}\dot{\phi^2}+V(\phi)\right]
\\
\frac{\ddot{\tilde{a}}}{\tilde{a}} &=&-\frac{1}{6M_p^2}\left[\tilde\rho(\phi)+3\tilde{P}(\phi)+2\dot{\tilde{\phi}}^2-2V(\phi)\right]
\eea
however they contain non-minimal couplings between the scalar and matter, which lead to the matter effectively feeling a fifth force. Matter test particles do not move along geodesics of the metric $\tilde{g}_{\mu\nu}$ and the density of matter is not purely affected by cosmic expansion but also by direct interconversion to $\phi$. This is reflected directly in the scalar and matter fluid equations,
\bea
\ddot{\phi}+3\tilde{H}\dot\phi+V,_{\phi} &=&\frac{1}{2M_p}\beta(\tilde\rho-3\tilde{P})
\\
\dot{\tilde{\rho}}+3\tilde{H}(\tilde\rho+\tilde{P})&=&-\frac{1}{2M_p}\beta(\tilde\rho-3\tilde{P})
\eea
where ${\tilde \rho}\equiv e^{-2 \beta\phi/M_p} \rho$ and ${\tilde P}\equiv e^{-2 \beta\phi/M_p} P$ are the Einstein frame energy density and pressure.

The Einstein and Jordan frames are wholly equivalent, neither one is more ``correct" than the other, however one has to think carefully about underlying assumptions made the interpretation of observations, for example how redshifts are interpreted, in order to match theoretical predictions to observational data. Typically redshifts are interpreted assuming minimally coupled matter, i.e. the Jordan frame.

Extensions of $f(R)$ gravity  have also been considered, that involve terms including scalar combinations of the Ricci tensor and Riemann tensor, $f(R,R^{\mu\nu}R_{\mu\nu},R^{\alpha\beta\mu\nu}R_{\alpha\beta\mu\nu})$  \cite{Carroll:2004de} . An example of this are models in which the action is modified by including the Gauss-Bonnet invariant,  $G\equiv R^2-4R^{\mu\nu}R_{\mu\nu}+R^{\alpha\beta\mu\nu}R_{\alpha\beta\mu\nu}$, \cite{Nojiri:2005jg}  chosen because it does not contain any badly behaved ``ghost-like" degrees of freedom. However it is difficult to find such $f(G)$ models that are consistent with current observational data \cite{Li:2007jm}.

{\bf The DGP model:} Another scenario is that gravity is not confined to 4D spacetime, and that it is able to pervade extra dimensions. One of the most widely studied extra dimension models is the Dvali-Gabadadze-Porratti (DGP) model \cite{Dvali:2000hr} whose cosmology is outlined in \cite{Lue:2005ya} . In this theory, matter is confined to a 4D ``brane" in a 5D flat, Minkowski ``bulk" of infinite volume. The cosmic acceleration we observe is then due to an effective 4D theory of gravity on the brane. 

The 5D action for DGP is
\bea
S = \int d^5x \sqrt{-g_{5}}\frac{M_5^3R_5}{2}+\int d^4x\sqrt{-g_4}\frac{M_4^2R_4}{2}\left[R_4 + {\mathcal L}_m\right]
\eea
where $g_n$, $M_n$ and $R_n$ are the metric, Planck mass and Ricci scalar in n dimensions.

The behavior of gravity in these models transitions at a cross-over scale $r_c$,
\bea
r_c= \frac{M_4^2}{2M_5^3}.
\eea
On smaller distances from the source, $r<r_c$, gravity appears four-dimensional, with $\Phi\propto 1/r$, while on large scales the graviton's interaction with the fifth dimension becomes important and  gravity is weakened, with $\Phi\propto 1/r^2$. Though gravity is four-dimensional on scales below $r_c$, it is different from standard General Relativity on distances down to the Vainshtein radius, $r_*\equiv (r_sr_c)^{1/3}$, where $r_s$ is the Schwarzschild radius for the mass.

The effective 4D Friedmann equation is
\bea
H^2+\frac{K}{a^2}-\epsilon\frac{1}{r_c}\sqrt{H^2+\frac{K}{a^2}}= \frac{8\pi G}{3}\rho
\eea
where $\epsilon=+1$ is the ``self-acceleration" branch of DGP, in which acceleration occurs when $H^{-1}\gtrsim r_c$, and $\epsilon=-1$ is the ``normal branch" in which one requires $\Lambda$ to generate acceleration. While acceleration can be generated naturally at late times in the self-accelerating branch for $r_c\sim H_0^{-1}$, it has been shown that the expansion history does not fit observational constraints from SNIa, BAO and the CMB as well as $\Lambda$CDM \cite{Rydbeck:2007gy,Fang:2008kc} . In addition, it is believed to suffer from the presence of a ghost-like degree of freedom \cite{Charmousis:2006pn}.

\subsubsection{k-essence}
Quintessence has a canonical kinetic energy term in its Langrangian ${\mathcal L}=X-V$ with $X=-\frac{1}{2}g^{\mu\nu}\partial_\mu\phi\partial_\nu\phi$, and relies on a slow evolution of the scalar field down a flat potential to drive late-time acceleration. An alternative scenario is that accleration is brought about because of a modification to the form of the kinetic Lagrangian, so-called {\bf k-essence} \cite{Chiba:1997ej,Chiba:1999ka,ArmendarizPicon:2000dh,ArmendarizPicon:2000ah}. This concept is a low-energy analogue of ``k-inflation" to explain inflation in the early universe \cite{Armendariz-Picon:1999rj}.

This generalized scalar field can be described by an action of the form
\bea
S&=& \int d^4x\sqrt{-g}{\mathcal L}(X,\phi)
\eea
which can contain terms including non-trivial functions of $X$ and combinations of $X$ and $\phi$. A possible motivation for actions of this form comes from considering low-energy effective actions from string theory in which higher-order derivative terms are important  \cite{Gasperini:2002bn}
\bea
S &=&\int d^4x\sqrt{-\tilde{g}}\left\{B_g(\phi)\tilde{R}+B_\phi^{(0)}(\phi)(\tilde\nabla\phi)^2-\alpha'\left[c_1^{(1)}B_\phi^{(1)}(\phi)(\tilde\nabla\phi)^4+...\right]+{\mathcal O}(\alpha')^2\right\}\hspace{1cm}
\eea
where $\phi$ is the dilaton field related to the string coupling by $g_s=e^{\phi}$, $\alpha'$ is related to the string length scale, $\lambda_s$, $\alpha'=\lambda_s/2\pi$. In the weak coupling regime $g_s\ll 1$, gives couplings $B_g\approx B_\phi^{(0)}\approx B_\phi^{(1)}\approx e^{-\phi}$. On doing a conformal transformation $g_{\mu\nu} = B_g(\phi)\tilde{g}_{\mu\nu}$ the model has an Einstein frame action that contains a k-essence style scalar Lagrangian
\bea
S&=& \int d^4x \sqrt{-g}\left[\frac{M_p^2}{2}R+K(\phi)X+L(\phi)X^2+...\right]
\\
K(\phi) &=& \frac{3}{2}\left(\frac{d\ln B_g}{d\phi}\right)^2-\frac{B_\phi^{(0)}}{B_g}
\\
L(\phi)&=& 2c_1^{(1)}\alpha'B_\phi^{(1)}(\phi).
\eea 
With a scalar field redefinition this transforms into a Lagrangian of the form
\bea
{\mathcal L}_\phi &=& f(\phi)(-X+X^2)
\eea
for which the equation of state is
\bea
w_\phi = \frac{1-X}{1-3X}
\eea
and acceleration is obtained for $X<2/3$.

In quintessence the minimum equation of state the scalar field can have is $w=-1$ when $X=0$, in contrast k-essence models can exhibit phantom like behavior with $w<-1$. Consider a general Lagrangian of the form ${\mathcal L} =K(X) - V(\phi)$, by varying the Lagrangian with respect to the metric the energy density and equation of state are given by
\bea
\rho &=& 2XK_X-K +V
\\
w &=& \frac{K-V}{2XK_X-K+V}=-1 +\frac{2XK_X}{2XK_X-K+V}
\eea
 where $K_X\equiv dK/dX$. For $K_X<0$ phantom behavior arises.
 
\subsubsection{Non-minimally coupled dark energy}

A theory in which the scalar field  is non-minimally coupled to matter can be described by the following action
\bea
S &=& S[g_{\mu\nu},\phi,\Psi_{\rm j}] \nonumber \\
&=& \int d^4x\sqrt{-g}
\left[ \frac{1}{2} M_{\rm p}^2 R
-\frac{1}{2} (\nabla \phi)^2
 - V(\phi)
\right] + \sum_j S_{\rm j}[e^{2 \alpha_{\rm j}(\phi)} g_{\mu\nu}, \Psi_{\rm j}], \hspace{0.5cm}
\label{action0}
\eea
where $g_{\mu\nu}$ is the Einstein frame metric, $\phi$ is a scalar field
which acts as dark energy, and $\Psi_{\rm j}$ are the matter fields.
The functions $\alpha_{\rm j}(\phi)$ are coupling functions that determine the strength
of the coupling of the jth matter sector to the scalar field.

The Friedmann equation and fluid equations for the scalar field and matter are
\bea
3 M_{p}^2 H^2 &=& \frac{1}{2} {\dot \phi}^2 + V(\phi) + \sum_{{\rm
    j}}  e^{\alpha_{j}(\phi)}{\rho}_{\rm j},\label{cons1}
\\
{\ddot \phi} + 3 H {\dot \phi} + V'(\phi) &=& - \sum_{\rm j} \alpha_{\rm
  j}'(\phi)  (1 - 3 w_{\rm j}) e^{\alpha_{j}(\phi)}{\rho}_{\rm j},
  \\
\dot {\rho}_{\rm j}+ 3 (1 + w_{\rm j}) H {  \rho}_{\rm j}
&=&  - 3  w_{\rm j} e^{\alpha_{j}(\phi)}{\rho}_{\rm j} \alpha_{\rm j}'(\phi) {\dot \phi},
\label{cons3}
\eea
where a dot represents a derivative with respect to the Einstein time coordinate, $t$, and primes are with respect to $\phi$, and $H = {\dot a}/a$ is the Einstein frame Hubble parameter.

We can solve equations (\ref{cons1})-(\ref{cons3}) in the scenario in which CDM alone is coupled to the scalar, by studying  Copeland et.\ al.'s \cite{Copeland:2006wr,Amendola:1999er} dimensionless quantities:
\begin{equation}
x\equiv \frac{{\dot \phi}}{\sqrt{6}H M_p} \ , \ \
y\equiv\frac{\sqrt{V}}{\sqrt{3}H M_p} \ , \ \ \lambda\equiv
-\frac{M_p V'}{V} \ , \ \ \Gamma\equiv \frac{VV''}{V'^2}.
\label{copelandvars}
\end{equation}
We also define a dimensionless variable to describe the coupling to matter,
\begin{equation}
C(\phi) \equiv -\frac{\alpha'(\phi) M_p}{\beta}.
\label{sigmadef}
\end{equation}
Rewriting in terms of the dependent variable
$N\equiv\ln(a)$ yields
\begin{eqnarray}
\frac{dx}{dN} &=& -3x+\frac{\sqrt{6}}{2}\lambda y^2
+\frac{3}{2}x(1+x^2-y^2) \nonumber \\
&&+C(1-x^2-y^2) \ , \\
\frac{dy}{dN} &=& -\frac{\sqrt{6}}{2}\lambda xy +\frac{3}{2}y(1+x^2-y^2) \ , \\
\frac{d\lambda}{dN} &=& -\sqrt{6}\lambda^2 (\Gamma -1)x.
\end{eqnarray}
In these equations, $\Gamma$, $C$ and $\alpha$ are understood to
be the functions of $\lambda$ obtained by eliminating $\phi$ in Eqs.\
(\ref{copelandvars}) and (\ref{sigmadef}).
After the equations have been solved, the matter density $\rho = \rho_c + \rho_b$ (neglecting radiation in the matter dominated era)
can be obtained from the equation
\begin{equation}
x^2+y^2+\frac{1}{3M_p^2H^2}\rho e^{\alpha}=1 \ .
\label{constraint}
\end{equation}
Note that the effective total equation of state parameter $w_{\rm eff}$, defined by
$a(t)\propto t^{2/3(1+w_{\rm eff})}$,
is simply given by
\begin{equation}
w_{\rm eff} = x^2-y^2 \ ,
\end{equation}
from Eq.\ (\ref{constraint}).

The fixed points of this system are the solutions of the equations
$dx/dN=dy/dN=d\lambda/dN=0$.

A specific example is provided by the exponential potential $V(\phi)=V_0 e^{-\lambda\phi/M_p}$
where $\lambda$ and $V_0$ are constants, and by $C(\phi)=C$, a constant.
The corresponding coupling function $\alpha(\phi)$ is then linear:
\be
\alpha(\phi) = -\beta C \phi/M_p.
\label{constantcoupling}
\ee

The fixed point followed in the matter dominated era is described by
\begin{eqnarray}
(x,y) &=& \left[\frac{2C}{3},0\right], \label{at1} 
\\
w_{\rm eff}  &=& \frac{4C^{2}}{9}\label{weff1}.
\eea
It doesn't give rise to acceleration but deviates from CDM evolution (with $w_{eff}=0$) depending on the coupling strength.

A second  attractor is independent of the non-minimal coupling and gives acceleration for $\lambda^2<2$
\bea
(x,y) &=& \left[ \frac{\lambda}{\sqrt{6}} \ , \ \left(1-\frac{\lambda^2}{6}\right)^{1/2}\right] \ , \label{at2}
\\
w_{\rm eff}&=&-1+\frac{\lambda^2}{3} \label{weff2}.
\end{eqnarray}

For a power law potential, $V\propto \phi^{-n}$, the CDM dominated era has the same attractor (\ref{at1}) while the accelerated era attractor is different
\bea
(x,y)&=& [0,1]
\\
w_{eff} &=& -1
\eea
independent of the coupling, $C$, and the precise power law form of the potential, $n$.

Constraints on the non-minimal coupling strength, $C$ in light of recent CMB, BAO, supernovae and galaxy distribution data are $|C|<0.13$  for an exponential potential and $-0.055<C<0.066$ for the power law potential  at the 95\% confidence level \cite{Bean:2008ac} .

\section{Using the growth of structure to understand the origin of cosmic acceleration}

As discussed in section \ref{distance}, cosmological constraints on the properties of cosmic acceleration have, to date, primarily come from the homogeneous background expansion, and associated geometrical distance measures of the CMB angular diameter distance, the BAO acoustic scale and supernovae luminosity distances. However a projected increase in the number, depth and breadth of large scale structure surveys offer the opportunity to also constrain the origin of acceleration by investigating its affect on the growth of cosmic structure (the growth of over-densities and peculiar velocities that seed galaxies and clusters of galaxies). By contrasting expansion history and growth of structure measurements one can investigate evidence for dark energy perturbations and modifications to gravity \cite{Bean:2003fb,Weller:2003hw,Hu:2004yd,Ishak:2005zs,Hannestad:2005ak,Knox:2006fh,Bertschinger:2006aw,Wang:2007fsa}. 

\subsection{Linear perturbation theory}

In order to study the growth of inhomogeneities we need to consider a perturbed, inhomogeneous metric and energy momentum tensor. A general perturbed FRW metric is given by

\bea
ds^2 = -(1+2AY)dt^2 - aBY_idx^idt +a^2(\delta_ij+2H_LY\delta_ij +2H_TY_{ij})dx^idx^j \hspace{0.5cm}
\eea
where $Y$,$Y_j$ and $Y_{ij}$ are unit scalar, vector and tensor perturbations respectively, and $A$,$B$, $H_L$ and $H_T$ are arbitrary time and space dependent scalar perturbations.

Though it is certainly possible to perform the analysis in this general ``gauge ready" approach it is somewhat time consuming. Instead we will use an intuitive gauge (choice of $A$,$B$,$H_L$ and $H_T$) known as the conformal Newtonian gauge in which $A=\Psi$, $B=H_T=0$, $H_L=\Phi$. In this gauge $\Phi$ reduces to the standard Newtonian potential (that obeys Poisson's equation) on scales much smaller than the horizon, 
and for $GR$, in the matter dominated and accelerated eras, $\Phi=-\Psi$.

As well as perturbing the gravitational metric, we also need to describe the perturbed RHS of Einstein's equation, the perturbed energy-momentum tensor
\bea
T^{0}_0 &=& -(\rho(t)+\delta\rho(x,t))=-\rho(t)(1+\delta(x,t))
\\
T^{0}_j &=& (\rho(t)+P(t))vY_j
\\
T^{i}_{j} &=& P(t)\delta^{i}_j+\delta P(x,t) \delta^{i}_{j}Y + \frac{3}{2}(\rho(t)+P(t))\sigma(x,y)Y^i_j
\eea
where $\delta$ is the fractional energy density, $v$ is the peculiar velocity, and $\delta P$ and $\sigma$ are the isotropic pressure and anisotropic shear stress perturbations respectively. The isotropic shear stress is sometimes rewritten in terms of the speed of sound of a wave propogating in that matter species
\bea
\delta P &=& c_s^2\delta\rho.
\eea

The perturbed Einstein equations relate the metric and matter peturbations
\bea
k^2\Phi &=& 4\pi G a^2 \sum_i \rho_i\Delta_i
\\
k^2(\Phi+\Psi) &=& -12\pi G a^2\sum_i(\rho_i+P_i)\sigma_i
\eea
The first equation is the Poisson equation with $\Delta_i \equiv (\delta_i+3H(1+w_i)\frac{v_i}{k})$ the density perturbation in the matter's rest frame. The second reflects that in regimes where relativistic species (and hence anisotropic shear stresses) are negligible then $\Phi=-\Psi$. 

The comobing Hubble length, $c/aH$, gives a measure of the region that is able to communicate at a given time. Comoving modes for which $k>aH$ are ``subhorizon" while $k<aH$ are ``superhorizon". 

Energy momentum conservation, ${T^{\mu\nu}}_{;\mu}=0$, for the inhomogeneous matter gives us the perturbed fluid equations. In terms of conformal time, $\dot{\delta}=d\delta/d\tau$, and $\hub=d\ln a/d\tau$, for cold dark matter
\bea
\dot\delta_c&=&-ikv_c-3\dot\Phi
\\
\dot v_c &=& -\hub v_c -ik\Psi.
\eea
Combined these give a second order differential equation in $\delta$,
\bea
\ddot\delta_c-\hub\dot\delta_c+3\hub\dot\Phi+3\ddot\Phi+k^2\Psi=0.
\eea
On subhorizon scales the time derivatives of the potentials are subdominant to the term in $k^2$
\bea
\ddot\delta_c+\hub\dot\delta_c-4\pi Ga^2\sum_i\rho_i\delta_i=0. \label{deltaevol}
\eea
During accelerated expansion the second term, a Hubble damping term, drives suppression of growth in $\delta_c$. 

For general matter
\bea
\dot\delta&=&-(1+w)(ikv+3\dot\Phi)-3\hub(c_s^2-w)\delta
\\
\dot v &=& -\hub(1-3w)v -\frac{\dot{w}}{1+w}v-\frac{c_s^2}{1+w}ik\delta-ik\Psi+ik\sigma.
\eea
which gives
\bea
\ddot\delta+\hub\dot\delta-(4\pi Ga^2\sum_i\rho_i\delta_i-c_s^2k^2\delta)=0.
\eea
If $c_s^2>0$ it reduces the effective strength of gravity on scales $k<a\hub/c_s$ and suppresses the growth of the overdensity. For $c_s^2<0$ catastrophic growth of the inhomogeneity can occur. This instability has been particularly discussed in the context of non-minimally coupled dark energy-dark matter models \cite{Afshordi:2005ym,Bean:2008ac} .

One might naively think that $c_s^2=c_a^2\equiv \dot{P}/\dot\rho$, where $c_a^2$ is the adiabatic sound speed
\bea
c_a^2 &=& w-\frac{\dot{w}}{3\hub(1+w)}.
\eea
This is the case for perfect fluids such as CDM, baryons and photons. For scalar field dark energy, however, this would imply that a negative equation of state would often lead to catastrophic collapse of dark energy over-densities, and the formation of dense `nuggets' of dark energy on subhorizon scales, something we do not observe. 

In general a fluid need not be perfect in this way, and the pressure perturbations are not wholly specified by the background expansion history. This leads to a disparity between $c_a^2$ and $c_s^2$  because of ``entropy perturbations,
\bea
\delta P_{en} &\equiv &(c_s^2-c_a^2)\delta.
\eea
For minimally coupled quintessence one finds $c_s^2=1$ irrespective of its equation of state. For a more general, k-essence model $c_s^2$ depends on the form of the kinetic action
\bea
c_s^2 = \frac{P_{,X}}{\rho_{,X}} &=& \frac{{\mathcal L}_{,X}}{{\mathcal L}_{,X}+2X{\mathcal L}_{,XX}}
\eea
which opens up the possibility of $c_s^2<0$ and an associated instability leading to uncontrolled growth on scales below $c_s/aH$.

In modified gravity models, the Einstein equations are modified while the fluid equations remain unchanged. A common parameterization introduces two scale and time-dependent functions, $G_{eff}(k,a)$ and $R(k,a)$, 
\bea
k^2\Phi &=&  4\pi G_{eff} a^2\sum_i\rho_i\Delta_i\label{EE000i}
\\
\Psi +R\Phi&=&  -   12\pi G_{eff} a^2 \sum_i \rho_i(1+w)\frac{\sigma_i}{k^2} \label{EEij}. \ \ \ \ \
\eea
The first describes a modified Poisson equation in which the gravitational potential responds differently to the presence of matter, with an effective Newton's constant, $G_{eff}$, while the second allows an inequality between the two gravitational potentials, even at late times when anisotropic shear stresses are negligible.  In the $DGP$  model\cite{Koyama:2005kd} discussed in \ref{modgrav}, 
\bea
G_{eff} &=& G\left(1-\frac{1}{3\beta}\right)
\\
R &=&\frac{\left(1+\frac{1}{3\beta}\right)}{\left(1-\frac{1}{3\beta}\right)}
\eea
where
\bea
\beta &\equiv & 1-2r_cH\left(1+\frac{\dot{H}}{3H^2}\right).
\eea
For $f(R)$ theories, $G_{eff}=G/(1+f_R)$ while $R$ is a more involved function obtained by integrating the Einstein and fluid equations \cite{Song:2006ej,Hu:2007pj} .

By measuring the evolution of the gravitational potentials, and their relationships to the growth of overdensities and peculiar velocities we can search for signatures of dark energy innhomogeneities, or modifications to gravity associated with cosmic acceleration. 

\subsection{Observations of large scale structure}

For a range of different observations, for example the Integrated Sachs Wolfe (ISW) effect in the CMB, galaxy number counts and weak gravitational lensing we discuss below, the measurements are based on the two-point correlation function between fluctuations.

Under the Limber approximation, the 2D angular power spectrum for the correlation between two fields, $X$ and $Y$, is
\bea
C^{XY}_l &= &\int_0^{\chi_{\infty}} \frac{d\chi}{\chi^2}W_X(\chi)W_Y(\chi)T_X(k_l, \chi)T_Y(k_l, \chi)\Delta_{R}^{2}(k_l), \label{cldef} \hspace{0.25cm} 
\eea
where $W_X$ is the window function which gives the redshift window for the field $X$, and $T_X$ is the transfer function that relates the inhomogeneity in $X$ today to its primordial value, and $\Delta_{\cal R}^2(k)$ is the dimensionless primordial spectrum of curvature fluctuations. $k_l \approx l/\chi$ where $l$ is the multipole moment (analogue of Fourier mode on the spherical sky) and $\chi$ is the comoving distance.

The statistical uncertainty in the power spectrum is given by
\bea
\Delta C_{l} &=& \sqrt{\frac{2}{(2l+1)f_{sky}}}\left[C_l + \frac{\sigma^2}{n_{eff}}\right]
\eea
where $f_{sky}$ is the fraction of the sky covered by a survey, $\sigma^2$ is the variance for a single measurement, and $n_{eff}$ is the effective number density per steradian of measurements used in the analysis. The first term in the square bracket represents ``cosmic variance", the intrinsic sample variance  coming from only having a finite number of independent sky samples to measure correlations. This term dominates on large scales where the number of independent samples is small. The second term is the statistical ``shot noise" error coming from intrinsic variations in the objects being measured and instrumental noise.

\subsubsection{Integrated Sachs Wolfe (ISW) effect}\label{isw}

In section \ref{distance} we discussed how the angular diameter distance to the last scattering surface of the CMB provides an important distance measure to constrain dark energy properties. The large scale CMB temperature correlations provide additional information that can also be used to constrain dark energy however. On scales below the sound horizon at last scattering, the physics of Thomson scattering and electron/ baryon interaction dominate the CMB temperature correlations leading to the acoustic peaks. On scales larger than the sound horizon only gravity and the cosmic expansion history determine the correlations.  Of particular interest in the context of dark energy is how the recent onset of cosmic acceleration affects the correlations in the CMB on the very largest scales observable, {\bf the Integrated Sachs Wolfe (ISW) effect}.  

 The ISW effect arises because of the effect of a time-changing gravitational potential on the CMB photon energy. As the photons stream to us today they traverse gravitational potential wells, gaining energy as the photons fall into the well, and losing it as they climb out. In the matter dominated era, solving (\ref{deltaevol}) one finds $\delta_c\propto a$, and hence from (\ref{EE000i}) that $\Phi=-\Psi$ is constant. This means that there would be no net change in the photon energy after moving through the constant gravitational well. During the period of accelerated expansion however the suppression of growth in the overdensities causes gravitational potential wells to decay so that photons gain more energy on their descent into the well than they lose on their climb out. This gain in energy is translated into a boost in large scale correlation amplitude in the CMB power spectrum.

The CMB correlation comes from integrating along the photon's history since last scattering so that all redshifts are sampled, in (\ref{cldef}) therefore $W_{ISW}=1$ and the transfer function is dependent on the rate of change of the gravitational potentials 
\bea
T_{ISW} = e^{-\tau_{reion}}\left(-\dot{\tilde\Phi}+\dot{\tilde\Psi}\right),
\eea
where  $\tilde{X}$ is the transfer function of $X$, normalized so that $X^2(k,\chi) = \tilde{X}^2 (k,\chi)\Delta_{\cal R}^2(k)$. $\tau_{reion}$ is the reionization optical depth, taking into account that some CMB photons are scattered by free electrons created when the universe was reionized by the first stars $z\approx 10$.

The ISW signal, being a large scale effect, is dominated by cosmic variance. It has been measured to cosmic variance limits by the WMAP survey \cite{Nolta:2008ih} .

\subsubsection{Weak gravitational lensing}

When light travels through a gravitational potential well its path is distorted much as it would be through an optical lens. There are two primary effects of the lensing: the image can be magnified and the image can be distorted, or undergo ``shear", which can be, in a simple way, thought of as a circular source having an elliptical image.  The lensing can be thought of as a remapping of the 2D angular distribution of the source's surface brightness, $f^s$, to the observed image, $f^{obs}$, through a ``distortion matrix", $A_{ij}$
\bea
f^{obs}(\theta_i) &=& f^{s}(A_{ij}\theta_j)
\eea 
with
\bea
A_{ij}\equiv \frac{\partial\theta_i}{\partial \theta_j} = \left(\begin{array}{cc}1-\kappa-\gamma_1 & -\gamma_2 \\ -\gamma_2 & 1-\kappa+\gamma_1\end{array}\right).
\eea
Here $\kappa$ is the {\bf convergence} which is related to the magnification, and $\gamma = \gamma_1+i\gamma_2$ is the {\bf shear}.

The convergence is given by
\bea
\kappa&=& -\frac{1}{2}\int_0^{\chi_s}k^2(\Phi+\Psi)W_{\kappa}(\chi,\chi_s)d\chi
\eea
where $\chi_s$ is the comoving distance to the source. The window and transfer function in (\ref{cldef}) are
\bea
T_{\kappa} &=&-\frac{k^2}{2}(\Phi+\Psi)
\\
W_{\kappa}(\chi,\chi_s) &=& \chi \int _0^{\chi_s}d\chi' n(\chi') \frac{\chi'-\chi}{\chi'}
\eea
where $n(\chi)$ is the number density distribution of lensing galaxies along the line of sight normalized such that $\int_0^{\chi_s} n(\chi)=1$.

Lensing is sensitive to both dark energy's effect on cosmic geometry, through the expansion history's effect on the window function, and  the growth of structure, sensitive to both the expansion history and the relation between gravity and matter inhomogeneities, through the transfer function.

Lensing surveys such as those from the Canada France Hawaii Legacy Telescope (CFHTLS)  \cite{Semboloni:2005ct,Fu:2007qq,Dore:2007jh} and the HST COSMOS survey \cite{Massey:2007gh,Lesgourgues:2007te} have provided the first applications of lensing data to test the nature of dark energy. Planned surveys such as the Dark Energy Survey (DES), a $\sim$5000 square degree survey with galaxy survey density of $\sim 15$ per square arcmin and mean redshift $\sim 0.7$, and the Large Synoptic Survey Telescope (LSST), which will cover $\sim 20,000$ square degrees of sky, with $\sim 30$ galaxies per square arcmin and mean redshift 1.2, hope to provide richer and higher precision lensing data with which to test the origins of cosmic acceleration.

Systematic errors in weak lensing measurements include intrinsic correlations (intrinsic alignments) in galaxy shapes, uncertainties/ errors in photometric redshift estimates of galaxies and errors in shape measurement arising from anistropy of the point spread function (PSF) of the detector, atmospheric distortions and pointing errors amongst others.

\subsubsection{Cross-correlation of large scale structure observations}

The ISW and lensing observations directly probe the properties of the gravitational potential. By contrast, measurements of the distribution of luminous matter provide only a proxy for the gravitational potential. To recover the distribution of all matter, both dark and luminous, one needs information about how the distributions of dark matter and luminous matter are related, through a potentially scale and redshift dependent {\bf bias}. However by using cross-correlation of  galaxy, ISW and lensing variables one can reduce the sensitivity to bias \cite{Zhang:2007nk}.  With regards to understanding cosmic acceleration it has been shown that cross-correlations could provide a powerful tool with which to distinguish between modified gravtiy and dark energy by measuring $G_{eff}/G$ and the relationship between the two Newtonian potentials $\Phi/\Psi$ \cite{Caldwell:2007cw,Zhang:2007nk,Amendola:2007rr,Jain:2007yk,Bertschinger:2008zb,Song:2008vm,Schmidt:2008hc,Zhang:2008ba,Zhao:2008bn,Zhao:2009fn,Guzik:2009cm} .

\section{Final remarks}

These lectures have discussed how gravity on cosmic scales relates the evolution of spacetime scale and curvature to the properties of matter contained within it. Current observations of geometric distances to Type 1a supernovae, baryon acoustic oscillations and the Cosmic Microwave Background last scattering surface have provided consistent, strong evidence that the universe is undergoing accelerated expansion. This is seemingly at discord with the a universe in which gravity is described by GR and matter has positive pressure. We have considered a spectrum of possible theoretical origins for this cosmic acceleration: a perfectly homogeneous energy source with negative pressure, a ``cosmological constant"; a near-homogeneous, potentially dynamically evolving new form of matter, `` dark energy"; or that gravity deviates from GR on cosmic scales, ``modified gravity". We have discussed how the combination, and comparison, of observations of both geometric distances and the growth of large scale structure, such as the ISW effect, weak gravitational lensing and galaxy number counts, could potentially give us an insight into which of the theoretical possibilities is the origin of cosmic acceleration. Upcoming surveys in next few years should provide excellent data with which to address this central issue facing cosmology and particle physics today.

\section{Further reading}

The following reviews are good resources for further reading on dark energy and cosmic acceleration:
\begin{itemize}
\item ``Dynamics of dark energy" \cite{Copeland:2006wr}, Copeland, Sami and Tsujikawa, 2006
\item ``Dark energy and the accelerating universe"\cite{Frieman:2008sn}, Frieman, Turner and Huterer, 2008 
\item ``Approaches to understanding cosmic acceleration" \cite{Silvestri:2009hh}, Silvestri and Trodden, 2009
\end{itemize}

\section*{Acknowledgments}

I would like to thank the organizers of TASI 2009, Scott Dodelson and Csaba Csaki, for the invitation to give this series of lectures and the TASI attendees for their enthusiastic questions throughout the week.

  \bibliographystyle{unsrt}
\bibliography{ref}
  
 \end{document}